\documentclass[fp,twocolumn]{jpsj3}
\makeatletter
\def\ext@figure{lof}
\def\ext@table{lot}
\makeatother
\usepackage[dvipdfmx]{graphicx}
\usepackage[caption=false]{subfig}
\usepackage{algorithm}
\usepackage{algpseudocode}
\usepackage{txfonts}
\usepackage{url}

\usepackage{xcolor}
\usepackage{latexsym}
\usepackage{amsmath,amssymb}
\usepackage{braket}
\usepackage{cite}
\usepackage{booktabs}
\usepackage{multirow}

\newcommand{\HVRP}{\mathcal{H}^{\mathrm{VRP}}}
\newcommand{\HVRPobj}{\mathcal{H}^{\mathrm{VRP}}_{\mathrm{obj}}}
\newcommand{\HVRPp}{{\mathcal{H}'}^{\mathrm{VRP}}}
\newcommand{\HVRPpobj}{{\mathcal{H}'}^{\mathrm{VRP}}_{\mathrm{obj}}}
\newcommand{\HVRPpp}{{\mathcal{H}''}^{\mathrm{VRP}}}
\newcommand{\HVRPppobj}{{\mathcal{H}''}^{\mathrm{VRP}}_{\mathrm{obj}}}
\newcommand{\HQMKP}{\mathcal{H}^{\mathrm{QMKP}}}
\newcommand{\HQMKPobj}{\mathcal{H}^{\mathrm{QMKP}}_{\mathrm{obj}}}

\title{Ising-Machine-Assisted Large Neighborhood Search with Flexibly Tunable Subproblem Size}

\author{Koshiro Fujimoto$^{1}$, Masashi Yamashita$^{1}$, and Shu Tanaka$^{1,2,3,4,5}$\thanks{shu.tanaka@keio.jp}}
\inst{
$^1$Graduate School of Science and Technology, Keio University, Kanagawa 223-8522, Japan \\
$^2$Department of Applied Physics and Physico-Informatics, Keio University, Kanagawa 223-8522, Japan \\
$^3$Keio University Sustainable Quantum Artificial Intelligence Center (KSQAIC), Keio University, Tokyo 108-8345, Japan \\
$^4$Human Biology-Microbiome-Quantum Research Center (WPI-Bio2Q), Keio University, Tokyo 108-8345, Japan \\
$^5$Green Computing System Research Organization, Waseda University, Tokyo 162-0042, Japan
} 

\abst{
Ising machines are heuristic solvers for combinatorial optimization, but their solution quality can degrade when large-scale constrained problems are solved directly. 
Ising-machine-assisted large neighborhood search (LNS) instead repeatedly updates a feasible current solution by solving smaller subproblems. 
An existing feasibility-preserving method for the vehicle routing problem (VRP) re-optimizes the entire routes of selected vehicles and thus cannot adjust the subproblem size finely. 
We propose LNS-VT, which introduces the number of consecutive steps re-optimized per vehicle as a parameter to control the subproblem size finely while preserving feasibility. 
For a 300-site, 5-vehicle VRP, its best setting reduced the objective value by approximately 10\% relative to the existing method after 100 iterations, and the appropriate setting changed with the current solution quality. 
Applying the same principle to the quadratic multiple knapsack problem, we confirmed that an appropriate subproblem size also exists, indicating that subproblem-size control is important in Ising-machine-assisted LNS.
}

\begin{document}
\maketitle

\section{Introduction}
\label{sec: introduction}

A combinatorial optimization problem is the task of selecting, from a discrete set of candidate solutions, a solution that minimizes or maximizes a prescribed objective function.
Such problems arise in a wide range of applications, including vehicle routing~\cite{laporte1992vehicle}, scheduling~\cite{hoitomt1993practical}, and portfolio optimization~\cite{chang2000heuristics}.
In many combinatorial optimization problems, the number of candidate solutions grows exponentially with the number of decision variables.
For instances of practical size, exhaustive search is therefore generally intractable.
Against this background, Ising machines have attracted attention as heuristic solvers for combinatorial optimization problems~\cite{mohseni2022ising,chakrabarti2022quantum}.

An Ising machine is a computing device that searches for low-energy states of an energy function expressed as an Ising model or in the equivalent quadratic unconstrained binary optimization (QUBO) form~\cite{lucas2014ising,tanaka2017quantum,tanahashi2019application,zaman2021pyqubo,glover2022quantum}.
If the target combinatorial optimization problem is formulated so that its optimal solutions correspond to the ground states, or minimizers, of the Ising or QUBO energy function, low-energy states obtained by an Ising machine are expected to provide high-quality candidate solutions to the original problem.
Depending on the implementation, the search is driven by thermal fluctuations~\cite{kirkpatrick1983optimization,johnson1989optimization,johnson1991optimization}, quantum fluctuations~\cite{kadowaki1998quantum,santoro2002theory,hauke2020perspectives,chakrabarti2022quantum}, or other dynamics~\cite{mcmahon2016fully,goto2019combinatorial}, often controlled by an annealing-like schedule.
Ising machines have been applied to a variety of combinatorial optimization problems, including vehicle routing~\cite{irie2019quantum,bao2021approach,kanai2024annealing,kawase2026parallelizable}, scheduling~\cite{zhang2022solving}, portfolio optimization~\cite{rosenberg2015solving,tatsumura2023real,hidaka2023correlation,takahashi2025effectiveness}, materials science~\cite{Harris2018SpinGlass, King2018Topological, Utimula2021Ionic, sampei2023quantum, couzinie2025machine}, and black-box optimization~\cite{kitai2020designing,tamura2026black,kikuchi2026factorization}.

However, when a large-scale constrained combinatorial optimization problem is solved with an Ising machine, the naive approach of directly inputting the entire original problem as a single QUBO can face two difficulties.
Hereafter, we refer to this approach as the naive method.
First, the number of binary variables in the QUBO increases with the problem size and may exceed the number of variables that a given Ising machine can handle~\cite{venturelli2015quantum,kikuchi2023hybrid}.
Even when the QUBO can be input, the enlarged search space can degrade the quality of the solutions obtained within a practical computation time.
Second, when a constrained problem is expressed as a QUBO, the constraints are often incorporated into the objective function as penalty terms.
The search space of the Ising machine then contains solutions that violate the constraints~\cite{takehara2019multiple}.
Therefore, a low-energy state is not necessarily a high-quality feasible solution of the original problem.
For large-scale constrained problems, it is thus important to reduce the number of input variables while retaining a way to construct feasible solutions of the original problem.

Several approaches have been proposed to utilize Ising machines for large-scale problems, including variable fixing~\cite{qbsolv, okada2019improving,irie2021hybrid,zhang2022solving,hattori2025advantages,hattori2025impact}, divide-and-conquer methods~\cite{chapuis2017finding,peng2025hybrid,kamishima2025construction,liu2026towards}, and iterative optimization~\cite{karimi2017boosting,karimi2017effective,liu2019modeling,kikuchi2023hybrid,fukada2024hybrid,ide2025extending,eder2026revisiting}.
Variable fixing reduces the number of variables input to the Ising machine by fixing a subset of the variables.
However, because the fixed variables are not re-optimized in a given computation, the explorable solution space depends strongly on the choice of fixed variables.
In addition, for constrained problems, constraints involving both fixed and free variables must be handled carefully.
Divide-and-conquer methods partition the original problem into several subproblems and solve them separately.
However, when constraints or objective-function terms couple variables belonging to different subproblems, independently obtained partial solutions may be difficult to recombine into a feasible or high-quality solution of the original problem.
Therefore, in problem reduction for constrained problems, it is important not only to make the subproblems small but also to design them so that feasible solutions of the original problem can be constructed from subproblem solutions.

From this viewpoint, we focus on large neighborhood search (LNS)~\cite{shaw1998using}, a type of iterative optimization.
LNS starts from a feasible initial solution and generates a new candidate solution by re-optimizing part of the current solution.
In each iteration, a subproblem corresponding to a neighborhood of the current solution is generated, and the corresponding part of the current solution is replaced with the solution obtained by solving the subproblem.
In this paper, we call this replacement operation reinsertion.

Classical heuristics such as greedy algorithms and tabu search have been used as subproblem solvers, and Ising machines have recently been employed for this purpose~\cite{nishimura2019item,zhang2022solving}.
When an Ising machine is used as the subproblem solver, LNS becomes a framework that iteratively solves QUBOs smaller than the full QUBO of the original problem.
In this paper, we say that a subproblem generation method preserves feasibility if, for any feasible current solution and any feasible solution of the subproblem, reinsertion produces a feasible solution of the original problem.
By using such a feasibility-preserving subproblem generation method, we can update the current solution while keeping it feasible throughout the iterations.
In this respect, LNS is a natural framework for applying Ising machines to large-scale constrained combinatorial optimization problems.

However, the performance of Ising-machine-assisted LNS depends strongly on the design of the subproblems.
When the subproblem is small, the Ising machine is more likely to find a good solution within the corresponding neighborhood, but the improvement obtained in a single iteration is limited.
When the subproblem is large, the neighborhood may contain candidate solutions with larger improvements, but the solution quality achieved by the Ising machine can deteriorate because the number of variables increases.
Therefore, in Ising-machine-assisted LNS, simply making the subproblem small is not the goal.
The central issue is to control the number of variables in the subproblem, while preserving feasibility, so that this number lies within a range in which the Ising machine can obtain high-quality solutions.

In a previous study~\cite{yamashita2026Geometric}, a structure-based subproblem generation method for LNS was proposed for the vehicle routing problem (VRP).
In this paper, we call this method LNS-V.
In LNS-V, several vehicles are selected from a feasible current solution, and the assignment and visiting order of the sites currently visited by the selected vehicles are re-optimized.
The vehicles that are not selected and the sites visited by them are fixed to the values of the current solution.
Consequently, reinserting any feasible subproblem solution into the current solution yields a solution that satisfies the constraints of the original problem.
Thus, LNS-V is a feasibility-preserving subproblem generation method.
Numerical experiments in the previous study showed that LNS-V produced higher-quality solutions than the naive method.

On the other hand, LNS-V cannot finely adjust the number of variables in the subproblem.
In LNS-V, the subproblem size is controlled by changing the number of vehicles included in the subproblem.
Because the entire route of each selected vehicle is re-optimized, changing the number of selected vehicles by one can cause a large change in the number of variables when each vehicle visits many sites.
As a result, there can be problem settings in which the subproblem size cannot be placed in the range where the Ising machine achieves high solution quality.
That is, although LNS-V preserves feasibility, the number of variables in the subproblem is not tunable with sufficient granularity.

In this study, we propose LNS-VT, a subproblem generation algorithm that controls the number of variables in the subproblem more flexibly while retaining the feasibility-preserving construction of LNS-V.
In LNS-VT, in addition to the number of vehicles included in the subproblem, the length of the contiguous route segment to be re-optimized for each selected vehicle, namely the number of consecutive steps, is introduced as a control parameter.
This additional parameter relaxes the dependence of the subproblem size on the number of steps in the original problem and enables finer control of the number of variables.
Moreover, by imposing constraints consistent with the original problem on the vehicles, sites, and route segments included in the subproblem, and by fixing the remaining part of the current solution, LNS-VT remains feasibility-preserving.

The contributions of this study are threefold.
First, for the VRP, we construct LNS-VT and show that it can control the number of variables in the subproblem more finely than LNS-V while preserving feasibility.
Second, through numerical experiments on a VRP with 300 sites and 5 vehicles, we show that LNS-VT can obtain higher-quality solutions than LNS-V and the naive method under the tested settings.
In particular, the best tested setting of LNS-VT reduces the objective function value by approximately 10\% relative to LNS-V after 100 iterations, and it reaches a solution quality comparable to the final quality of LNS-V with approximately 30\% of the iterations.
Furthermore, we observe that the appropriate subproblem setting changes with the quality of the current solution.
Third, for the quadratic multiple knapsack problem (QMKP), we construct a subproblem generation method based on the same design principle and examine the relation between the number of variables in the subproblem and the solution quality.
The numerical results show that LNS with this method also obtains higher-quality solutions than the naive method for the QMKP in the tested settings and that the appropriate range of the number of variables differs substantially from that for the VRP.

These results suggest that, in Ising-machine-assisted LNS, preserving feasibility alone is not sufficient for obtaining high-quality solutions.
They also indicate that the ability to adjust the number of variables in the subproblem independently of the problem size and structure is important.
The similar dependence of solution quality on subproblem size observed for the VRP and the QMKP suggests that structure-aware and flexibly tunable subproblem generation can be useful for constrained combinatorial optimization problems.
At the same time, because the appropriate subproblem size depends on the problem class and on the quality of the current solution, subproblem generation and parameter selection should be designed with these dependencies in mind.

The rest of this paper is organized as follows.
In Sec.~\ref{sec: problem}, we define the VRP, the main problem treated in this study, and its QUBO formulation.
In Sec.~\ref{sec: method}, we describe the framework of Ising-machine-assisted LNS, the existing method LNS-V, and the proposed method LNS-VT.
In Sec.~\ref{sec: result}, we present the numerical calculation conditions and the results for the VRP.
In Sec.~\ref{sec: discussion}, we discuss the effects of the number of variables in the subproblem, the number of selected vehicles, and the quality of the current solution on the performance of LNS.
In Sec.~\ref{sec: QMKP}, we present the subproblem generation method and numerical results for the QMKP.
Finally, in Sec.~\ref{sec: conclusion}, we summarize this study and discuss future work.

\section{Problem}
\label{sec: problem}

In this section, we define the VRP, the main problem treated in this study, and give its QUBO formulation.
To isolate the effect of subproblem generation methods in LNS, we consider a single-depot uncapacitated VRP, in which each vehicle starts from and returns to the depot.
We distinguish between the objective function of the original VRP and the energy function of the QUBO input to the Ising machine.
In the QUBO formulation below, constraints A, B, and C are incorporated into the energy function as penalty terms that are zero when the corresponding constraint is satisfied and positive when it is violated.
Constraint D is imposed by fixing the corresponding variables before the problem is input to the Ising machine.

Let $N$ be the number of non-depot sites and $V$ be the number of vehicles.
The depot is labeled as site $0$, and the non-depot sites are indexed from $1$ to $N$.
We denote the set of non-depot sites by $\mathcal{N}=\{1,2,\ldots,N\}$ and the set of sites including the depot by $\mathcal{N}_0=\{0,1,\ldots,N\}$.
We also denote the set of vehicles by $\mathcal{V}=\{0,1,\ldots,V-1\}$.
Let $l_{i,j}$ denote the distance from site $i\in\mathcal{N}_0$ to site $j\in\mathcal{N}_0$.
We assume a symmetric distance matrix satisfying $l_{i,i}=0$ and $l_{i,j}=l_{j,i}>0$ for $i\neq j$.

The route of each vehicle is represented as a sequence of $T$ discrete steps.
We denote the set of steps by $\mathcal{T}=\{0,1,\ldots,T-1\}$ and assume that each vehicle is located at exactly one site at each step.
Each vehicle starts at the depot at step $0$ and is required to be at the depot at step $T-1$.
It may visit non-depot sites only at the intermediate steps $t=1,\ldots,T-2$.
Therefore, a single vehicle can visit at most $T-2$ non-depot sites.
Because the vehicles provide $V(T-2)$ intermediate steps in total, a necessary condition for all non-depot sites to be visited is
$V(T-2)\geq N$, or equivalently $T\geq\lceil N/V\rceil+2$ for integer $T$.
Here, $\lceil \cdot \rceil$ represents the ceiling function.
When a vehicle visits fewer than $T-2$ non-depot sites, the remaining intermediate steps after its return to the depot are represented as stays at the depot.

To formulate the VRP as a QUBO, we introduce the following binary variables $x_{t,i}^{(v)}$.
\begin{equation}
    \label{eq: VRP_input}
    x_{t,i}^{(v)} =
    \begin{cases}
        1, & \text{if vehicle $v$ visits site $i$ at step $t$},\\
        0, & \text{otherwise}.
    \end{cases}
\end{equation}
Here, $v\in\mathcal{V}$ is the vehicle index, $t\in\mathcal{T}$ is the step index, and $i\in\mathcal{N}_0$ is the site index.
The full formulation introduces $V\times T\times (N+1)$ binary variables.
However, as described below, constraint D fixes all variables corresponding to the first and last steps.
After the fixed values are substituted, the QUBO input to the Ising machine in the naive method contains $V(T-2)(N+1)$ free binary variables.

The objective function of the original VRP is the total travel distance of all vehicles.
In terms of the binary variables $x_{t,i}^{(v)}$, the total travel distance is given by
\begin{align}
    \label{eq: VRPobjective}
    \HVRPobj(\boldsymbol{x})
    =
    \sum_{v\in\mathcal{V}}
    \sum_{t=0}^{T-2}
    \sum_{i\in\mathcal{N}_0}
    \sum_{j\in\mathcal{N}_0}
    l_{i,j}x_{t,i}^{(v)}x_{t+1,j}^{(v)}.
\end{align}
For each vehicle, this expression sums the distances between the sites visited at consecutive steps.
The goal of the VRP considered in this study is to minimize $\HVRPobj(\boldsymbol{x})$ subject to the following constraints.

\begin{description}
    \item[Constraint A]
    Each non-depot site is visited exactly once by one of the vehicles.
    \begin{align}
        \label{eq: VRPconstA}
        \sum_{v\in\mathcal{V}}\sum_{t\in\mathcal{T}}x_{t,i}^{(v)}=1,
        \quad \forall i\in\mathcal{N}.
    \end{align}

    \item[Constraint B]
    Each vehicle visits exactly one site at each step.
    \begin{align}
        \label{eq: VRPconstB}
        \sum_{i\in\mathcal{N}_0}x_{t,i}^{(v)}=1,
        \quad \forall v\in\mathcal{V},\ \forall t\in\mathcal{T}.
    \end{align}

    \item[Constraint C]
    If a vehicle is at the depot at any step $t\geq 1$, it remains at the depot at the next step.
    \begin{align}
        \label{eq: VRPconstC}
        x_{t,0}^{(v)}=1
        \Rightarrow
        x_{t+1,0}^{(v)}=1,
        \quad
        \forall v\in\mathcal{V},\ \forall t\in\{1,2,\ldots,T-2\}.
    \end{align}

    \item[Constraint D]
    Every vehicle starts from and ends at the depot.
    \begin{gather}
        x^{(v)}_{0,0}=1,\quad \forall v\in\mathcal{V},
        \label{eq: VRPconstD1}\\
        x^{(v)}_{0,i}=0,\quad \forall v\in\mathcal{V},\ \forall i\in\mathcal{N},
        \label{eq: VRPconstD2}\\
        x^{(v)}_{T-1,0}=1,\quad \forall v\in\mathcal{V},
        \label{eq: VRPconstD3}\\
        x^{(v)}_{T-1,i}=0,\quad \forall v\in\mathcal{V},\ \forall i\in\mathcal{N}.
        \label{eq: VRPconstD4}
    \end{gather}
\end{description}

Constraint A is a global constraint that enforces the uniqueness of visits across all vehicles.
In contrast, constraints B, C, and D enforce the consistency of each vehicle route.
This distinction is used in the LNS subproblem generation described in Sec.~\ref{sec: method}, where both the uniqueness of visits and the route consistency must be preserved after reinsertion.

Constraint D is enforced by variable fixing rather than by penalty terms.
After the fixed values specified by constraint D are substituted, constraints A, B, and C are incorporated as penalty terms.
For compactness, the following expression is written using the original index sets.
Terms containing only fixed variables are constants and can be removed from the actual QUBO input to the Ising machine.
The resulting QUBO energy function is
\begin{align}
    \label{eq: VRPenergy}
    \HVRP(\boldsymbol{x})
    =
    \HVRPobj(\boldsymbol{x})
    &+
    \mu_\mathrm{A}
    \sum_{i\in\mathcal{N}}
    \left(
        \sum_{v\in\mathcal{V}}\sum_{t\in\mathcal{T}}x_{t,i}^{(v)}-1
    \right)^2
    \nonumber\\
    &+
    \mu_\mathrm{B}
    \sum_{v\in\mathcal{V}}\sum_{t\in\mathcal{T}}
    \left(
        \sum_{i\in\mathcal{N}_0}x_{t,i}^{(v)}-1
    \right)^2
    \nonumber\\
    &+
    \mu_\mathrm{C}
    \sum_{v\in\mathcal{V}}\sum_{t=1}^{T-2}
    x_{t,0}^{(v)}
    \left(1-x_{t+1,0}^{(v)}\right).
\end{align}
Here, $\mu_\mathrm{A}$, $\mu_\mathrm{B}$, and $\mu_\mathrm{C}$ are positive penalty coefficients corresponding to constraints A, B, and C, respectively.
Constraints A and B are one-hot constraints, and the first and second penalty terms are the corresponding standard quadratic penalties~\cite{lucas2014ising}.
The first penalty term is positive when a non-depot site is not visited exactly once.
The second penalty term is positive when the one-site-per-step condition is violated.
Constraint C is equivalent to the monotonicity condition $x_{t,0}^{(v)}\leq x_{t+1,0}^{(v)}$ for $t\in\{1,\ldots,T-2\}$ and thus acts as a domain-wall constraint on the depot variables of each vehicle~\cite{chancellor2019domain}.
The third penalty term is positive when a vehicle that is at the depot moves to a non-depot site at the next step, namely, when this monotonicity is violated.

The index $t=0$ is excluded from constraint C and from the third penalty term.
If $t=0$ were included, then, because every vehicle is at the depot at step $0$ owing to constraint D, no vehicle could depart from the depot.
Thus, constraint C is imposed only from $t=1$ onward.

After expansion, Eq.~\eqref{eq: VRPenergy} consists only of constant, linear, and quadratic terms in the binary variables.
Because $(x_{t,i}^{(v)})^2=x_{t,i}^{(v)}$, the linear terms can be represented as diagonal terms in the QUBO matrix.
Therefore, Eq.~\eqref{eq: VRPenergy} gives a QUBO formulation of the VRP.
If the penalty coefficients are chosen sufficiently large relative to the possible variation of the objective function, the minimizers of Eq.~\eqref{eq: VRPenergy} coincide with the optimal feasible solutions of the original VRP~\cite{lucas2014ising}.

\section{Method}
\label{sec: method}

In this section, we describe the framework of Ising-machine-assisted LNS, in which an Ising machine is used as the subproblem solver, and the subproblem generation methods for the VRP.
In Sec.~\ref{subsec:LNS}, we describe the update procedure of LNS used in this study.
In Sec.~\ref{subsec: LNS-V}, we define the existing method LNS-V and summarize its feasibility preservation and its limitation on subproblem-size control.
In Sec.~\ref{subsec: LNS-VT}, we define the proposed method LNS-VT and clarify its differences from LNS-V.

\subsection{Ising-machine-assisted LNS}
\label{subsec:LNS}

LNS is an iterative search method that generates a new candidate solution by re-optimizing part of a feasible current solution.
In the following, we consider a minimization problem.
Maximization problems can be treated by reversing the sign of the objective function.
We denote the objective function of the original problem by $f(\boldsymbol{x})$ and the feasible region by $\mathcal{F}$, and write the current solution at the $m$-th iteration as $\boldsymbol{x}^{(m)}\in\mathcal{F}$.

In simple local search, the solutions in a neighborhood of the current solution are often enumerated, and an improving solution is sought among them.
The computational cost of such an enumeration increases rapidly as the neighborhood becomes large.
In contrast, LNS defines a large neighborhood as a subproblem by making part of the current solution variable and fixing the remaining part.
The large neighborhood of a current solution $\boldsymbol{x}$ is denoted by $\mathcal{LN}(\boldsymbol{x})$.
In LNS, the search within this neighborhood is performed not by exhaustive enumeration but by a heuristic or metaheuristic solver.

In this study, we use an Ising machine as the subproblem solver of LNS.
That is, in each iteration, a QUBO smaller than the full QUBO of the original problem is generated and solved with the Ising machine.
The obtained subproblem solution is reinserted into the corresponding part of the current solution and gives a candidate solution $\boldsymbol{x}'$ of the original problem.
As described in Sec.~\ref{sec: introduction}, this operation of replacing part of the current solution with the subproblem solution is called reinsertion.

The outline of the LNS procedure used in this study is shown in Algorithm~\ref{alg:LNS}.
In the following, a solution that satisfies all the constraints of the subproblem is called subproblem-feasible.
When multiple solution candidates are obtained by solving the subproblem, the subproblem-feasible solution that yields the smallest objective function value after reinsertion is adopted.
If no subproblem-feasible solution is obtained, the current solution is kept unchanged.
The current solution is updated only when the candidate solution $\boldsymbol{x}'$ obtained by reinsertion satisfies the constraints of the original problem and improves the objective function value.

\begin{algorithm}[t]
  \caption{Large Neighborhood Search (LNS)}
  \label{alg:LNS}
  \begin{algorithmic}[1]
    \Require the number of iterations $M$
    \Ensure approximate solution $\boldsymbol{x}^*$
    \State Prepare a feasible initial solution $\boldsymbol{x}^{(0)}\in\mathcal{F}$
    \For{$m = 0$ \textbf{to} $M-1$}
      \State Generate a subproblem corresponding to the large neighborhood $\mathcal{LN}(\boldsymbol{x}^{(m)})$ of the current solution $\boldsymbol{x}^{(m)}$
      \State Solve the subproblem with the Ising machine
      \If{a subproblem-feasible solution is obtained}
        \State Reinsert the best subproblem-feasible solution into $\boldsymbol{x}^{(m)}$ to generate a candidate solution $\boldsymbol{x}'$
        \If{$\boldsymbol{x}'\in\mathcal{F}$ and $f(\boldsymbol{x}') < f(\boldsymbol{x}^{(m)})$}
          \State $\boldsymbol{x}^{(m+1)} \leftarrow \boldsymbol{x}'$
        \Else
          \State $\boldsymbol{x}^{(m+1)} \leftarrow \boldsymbol{x}^{(m)}$
        \EndIf
      \Else
        \State $\boldsymbol{x}^{(m+1)} \leftarrow \boldsymbol{x}^{(m)}$
      \EndIf
    \EndFor
    \State $\boldsymbol{x}^* \leftarrow \boldsymbol{x}^{(M)}$
    \State \Return $\boldsymbol{x}^*$
  \end{algorithmic}
\end{algorithm}

When LNS is applied to constrained combinatorial optimization problems, the subproblem generation method should satisfy two design requirements in the present framework.
The first requirement is feasibility preservation.
As defined in Sec.~\ref{sec: introduction}, a subproblem generation method preserves feasibility if, for any feasible current solution and any feasible subproblem solution, reinsertion produces a feasible solution of the original problem.
The second requirement is that the number of variables in the subproblem can be adjusted to a range in which the Ising machine can obtain high-quality solutions.
If the subproblem is too small, the improvement obtained in a single iteration is limited.
If the subproblem is too large, the solution quality returned by the Ising machine can deteriorate because the search space becomes larger, and the current solution may not be improved.
Therefore, Ising-machine-assisted LNS requires a subproblem generation method that preserves feasibility and allows the number of variables in the subproblem to be adjusted appropriately.
In addition to these requirements on the subproblem generation method, the framework assumes that a feasible initial solution is available.

\subsection{Existing subproblem generation method LNS-V}
\label{subsec: LNS-V}

In this subsection, we summarize the subproblem generation method proposed in the previous study~\cite{yamashita2026Geometric}, which exploits the problem structure of the VRP.
In this paper, we call this method LNS-V.
LNS-V selects several vehicles from the current solution and re-optimizes the assignment and visiting order of the non-depot sites visited by them.
The vehicles that are not selected and the non-depot sites visited by them are fixed to the values of the current solution.
With this construction, reinserting any feasible subproblem solution of LNS-V into the current solution yields a feasible solution of the original problem.

In the following, we consider a feasible current solution $\boldsymbol{x}^{(m)}$ and write its components as $x_{t,i}^{(v,m)}$.
We select $V'$ vehicles uniformly at random without replacement from the vehicle set $\mathcal{V}$ and denote the set of the selected vehicles by $\mathcal{V}_{\mathrm{sub}}$.
In this study, we treat the case of $2\leq V'\leq V$ because we focus on updates that can exchange visited sites between vehicles.
Although the case of $V'=1$ can also be defined, it is excluded from the comparison in this study because no exchange of visited sites between vehicles occurs in that case.

\begin{figure}[t]
  \centering
  \subfloat[]{\includegraphics[width=0.40\linewidth]{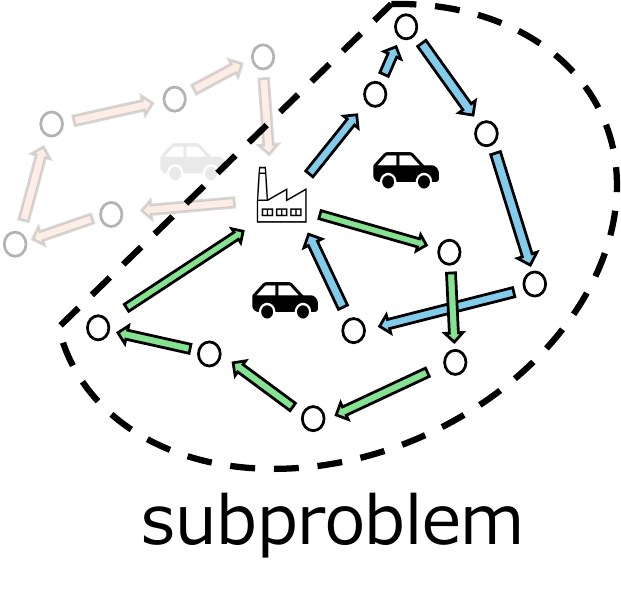}}
  \hspace{0.10\linewidth}
  \subfloat[]{\includegraphics[width=0.40\linewidth]{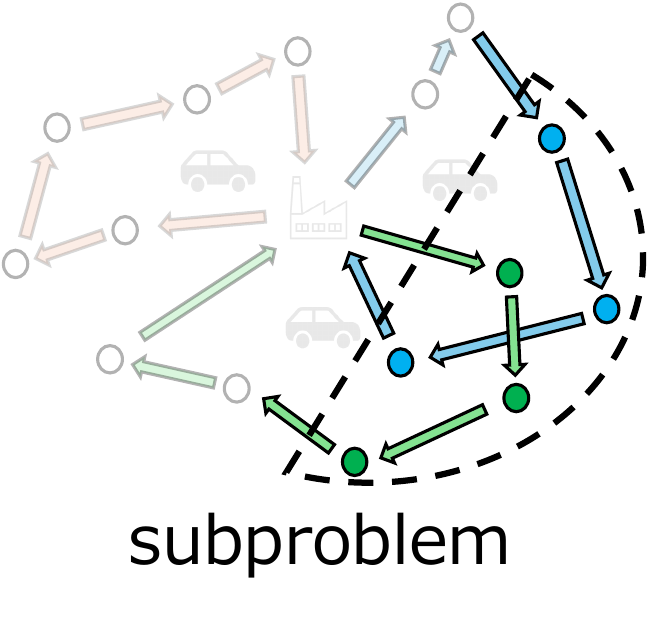}}
  \caption{Schematics of the vehicles and sites included in the subproblems of LNS-V and LNS-VT.
  (a) Schematic of LNS-V, where the entire routes of the selected vehicles are re-optimized.
  (b) Schematic of LNS-VT, where only a contiguous part of the steps in the route of each selected vehicle is re-optimized.}
  \label{fig:VRP_sub}
\end{figure}

The set of non-depot sites visited by the selected vehicles in the current solution is defined as
\begin{align}
    \mathcal{N}_{\mathrm{sub}}
    =
    \left\{
        i\in\mathcal{N}
        \mid
        \exists v\in\mathcal{V}_{\mathrm{sub}},\ 
        \exists t\in\{1,\ldots,T-2\},\
        x_{t,i}^{(v,m)}=1
    \right\}.
\end{align}
We also define the site set of the subproblem including the depot as $\mathcal{N}_{\mathrm{sub},0}=\mathcal{N}_{\mathrm{sub}}\cup\{0\}$.
The free variables of the subproblem are $x_{t,i}^{(v)}$ with $v\in\mathcal{V}_{\mathrm{sub}}$, $t\in\{1,\ldots,T-2\}$, and $i\in\mathcal{N}_{\mathrm{sub},0}$.
All the other variables are fixed to the values of the current solution.
In the subproblem of LNS-V, we re-optimize which vehicle in $\mathcal{V}_{\mathrm{sub}}$ visits each non-depot site in $\mathcal{N}_{\mathrm{sub}}$ and in which order.
Figure~\ref{fig:VRP_sub}(a) shows a schematic of the subproblem in LNS-V.

The objective function of the subproblem in LNS-V is the part of the total travel distance that depends on the selected vehicles.
After the fixed variables are substituted, this objective function can be written compactly as
\begin{align}
    \label{eq: lns_v_obj}
    \HVRPpobj(\boldsymbol{x})
    =
    \sum_{v\in\mathcal{V}_{\mathrm{sub}}}
    \sum_{t=0}^{T-2}
    \sum_{i\in\mathcal{N}_{\mathrm{sub},0}}
    \sum_{j\in\mathcal{N}_{\mathrm{sub},0}}
    l_{i,j}x_{t,i}^{(v)}x_{t+1,j}^{(v)}.
\end{align}
Here, the variables corresponding to the depot visits at the first and last steps are fixed according to constraint D described in Sec.~\ref{sec: problem}.
Terms that contain only fixed variables are constants and can be omitted from the actual QUBO input to the Ising machine.

Next, we consider the constraints of LNS-V.
As described in Sec.~\ref{sec: problem}, constraint A of the VRP is a global constraint that represents the uniqueness of visits across all vehicles.
In contrast, constraints B, C, and D represent the consistency of each vehicle route.
In LNS-V, the variables corresponding to the vehicles that are not selected are fixed to the values of the current solution.
Therefore, constraints B, C, and D need to be imposed only on the selected vehicles.
In addition, the variables of the selected vehicles at the non-depot sites outside $\mathcal{N}_{\mathrm{sub}}$ are fixed to zero.
Hence, the numbers of visits to the non-depot sites outside $\mathcal{N}_{\mathrm{sub}}$ are not changed by the subproblem solution, and for constraint A, it is sufficient to impose that each non-depot site in $\mathcal{N}_{\mathrm{sub}}$ is visited exactly once by the selected vehicles.
For the same reason, constraint B for the selected vehicles can be written using only the sites in $\mathcal{N}_{\mathrm{sub},0}$.

From the above, the QUBO energy function for the subproblem of LNS-V is given by
\begin{align}
    \label{eq: lns_v_energy}
    \HVRPp(\boldsymbol{x})
    =
    \HVRPpobj(\boldsymbol{x})
    &+
    \mu_\mathrm{A}
    \sum_{i\in\mathcal{N}_{\mathrm{sub}}}
    \left(
        \sum_{v\in\mathcal{V}_{\mathrm{sub}}}
        \sum_{t=1}^{T-2}
        x_{t,i}^{(v)}
        -1
    \right)^2
    \nonumber\\
    &+
    \mu_\mathrm{B}
    \sum_{v\in\mathcal{V}_{\mathrm{sub}}}
    \sum_{t=1}^{T-2}
    \left(
        \sum_{i\in\mathcal{N}_{\mathrm{sub},0}}
        x_{t,i}^{(v)}
        -1
    \right)^2
    \nonumber\\
    &+
    \mu_\mathrm{C}
    \sum_{v\in\mathcal{V}_{\mathrm{sub}}}
    \sum_{t=1}^{T-2}
    x_{t,0}^{(v)}
    \left(1-x_{t+1,0}^{(v)}\right).
\end{align}
The first penalty term penalizes violations of the condition that each non-depot site included in the subproblem is visited exactly once.
The second penalty term penalizes violations of the condition that each selected vehicle visits exactly one site at each intermediate step.
The third penalty term penalizes the case in which a selected vehicle that is at the depot moves to a non-depot site at the next step.

This construction preserves feasibility in the following sense.
First, the variables associated with the non-depot sites not included in $\mathcal{N}_{\mathrm{sub}}$ are all fixed to the values of the current solution, and hence each of these sites remains visited exactly once.
For any feasible solution of the subproblem, the first penalty term in Eq.~\eqref{eq: lns_v_energy} is zero, and hence each non-depot site in $\mathcal{N}_{\mathrm{sub}}$ is visited exactly once by one of the selected vehicles.
Therefore, constraint A of the original problem remains satisfied after reinsertion.
In addition, for any feasible solution of the subproblem, the second and third penalty terms in Eq.~\eqref{eq: lns_v_energy} are zero, and the fixed depot conditions at the first and last steps are satisfied.
Thus, the routes of the selected vehicles satisfy constraints B, C, and D.
For the vehicles that are not selected, the current solution is kept unchanged, and constraints B, C, and D are inherited.
Consequently, reinserting any feasible subproblem solution of LNS-V into the current solution yields a feasible solution of the original problem.

On the other hand, in LNS-V, the number of variables in the subproblem cannot be adjusted finely.
In LNS-V, the subproblem size is controlled by changing the number of selected vehicles $V'$.
Because the entire route of each selected vehicle is re-optimized, the number of steps in the subproblem is fixed to the number of steps $T$ of the original problem.
The number of intermediate steps is $T-2$, and the number of non-depot sites included in the subproblem is at most $V'(T-2)$.

If all free variables including the depot variables are counted, the total number of free binary variables in LNS-V is
\begin{align}
    N_{\mathrm{free}}^{\mathrm{V}}
    =
    V'(T-2)\left(|\mathcal{N}_{\mathrm{sub}}|+1\right).
\end{align}
The number of free variables corresponding to non-depot sites is
\begin{align}
    \label{eq: lns_v_variable_number_general}
    n_{\mathrm{var}}^{\mathrm{V}}
    =
    V'(T-2)|\mathcal{N}_{\mathrm{sub}}|
    \leq
    (V')^2(T-2)^2.
\end{align}
The difference between $N_{\mathrm{free}}^{\mathrm{V}}$ and $n_{\mathrm{var}}^{\mathrm{V}}$ is the number $V'(T-2)$ of free depot variables.
In the following discussion of subproblem-size scaling, we represent the subproblem size by $n_{\mathrm{var}}^{\mathrm{V}}$ because the site-assignment variables dominate.
In particular, when each selected vehicle visits $T-2$ non-depot sites, we have $|\mathcal{N}_{\mathrm{sub}}|=V'(T-2)$ and
\begin{align}
    \label{eq: lns_v_variable_number}
    n_{\mathrm{var}}^{\mathrm{V}}
    =
    (V')^2(T-2)^2.
\end{align}
Equation~\eqref{eq: lns_v_variable_number} shows that the number of site-assignment variables in the subproblem of LNS-V depends strongly on $V'$ and on the number of steps $T$ of the original problem.
Therefore, in problem settings with large $T$, changing $V'$ by one can cause a large change in the number of variables in the subproblem.
This limited granularity makes it difficult to place the subproblem size in a range where the Ising machine can obtain high-quality solutions.

\subsection{Proposed subproblem generation method LNS-VT}
\label{subsec: LNS-VT}

As described in the previous subsection, LNS-V preserves feasibility, but it cannot finely adjust the number of variables in the subproblem.
In this study, we therefore introduce the prescribed number of consecutive steps to be re-optimized, denoted by $T'$, in addition to the number of selected vehicles $V'$.
We call this subproblem generation method LNS-VT.
LNS-VT is designed to control the number of variables in the subproblem more finely while retaining the feasibility-preserving construction of LNS-V.

For each vehicle $v$, we first define the number of steps at which vehicle $v$ visits non-depot sites in the current solution as
\begin{align}
    T_{\mathrm{valid}}^{(v)}
    =
    \sum_{t=1}^{T-2}\sum_{i\in\mathcal{N}}x_{t,i}^{(v,m)}.
\end{align}
Owing to constraint C, once a vehicle is at the depot after step $0$, it remains at the depot.
Thus, the steps at which vehicle $v$ visits non-depot sites appear consecutively as $1,\ldots,T_{\mathrm{valid}}^{(v)}$.
We define the set of vehicles that visit at least one non-depot site as
\begin{align}
    \mathcal{V}_{+}^{(m)}
    =
    \left\{
        v\in\mathcal{V}
        \mid
        T_{\mathrm{valid}}^{(v)}\geq 1
    \right\}.
\end{align}
We assume that $|\mathcal{V}_{+}^{(m)}|\geq V'$ in the following.
If this condition is not satisfied in an implementation, the iteration can be skipped or $V'$ can be reduced.

We select $V'\ (2\leq V'\leq V)$ vehicles uniformly at random without replacement from $\mathcal{V}_{+}^{(m)}$ and denote the set of selected vehicles by $\mathcal{V}_{\mathrm{sub}}$.
The prescribed segment length $T'$ may be larger than the number of non-depot visits of some selected vehicle.
Therefore, for the generated subproblem, we define the actual segment length as
\begin{align}
    T_{\mathrm{seg}}
    =
    \min\left(
        T',\ 
        \min_{v\in\mathcal{V}_{\mathrm{sub}}}T_{\mathrm{valid}}^{(v)}
    \right).
\end{align}
Then, for each selected vehicle $v\in\mathcal{V}_{\mathrm{sub}}$, we extract $T_{\mathrm{seg}}$ consecutive steps from the non-depot part of its route.
Here, we write the site visited by vehicle $v$ at step $t$ in the current solution as $r_v^{(m)}(t)$.
Because the extraction interval is chosen only from the non-depot part of the route, every site visited at a step in the interval is a non-depot site.

For each vehicle $v\in\mathcal{V}_{\mathrm{sub}}$, the extraction start step $T_0^{(v)}$ is chosen uniformly at random from the range satisfying
\begin{align}
    \label{eq: lnsvt_start_condition}
    1\leq T_0^{(v)}
    \leq
    T_{\mathrm{valid}}^{(v)}-T_{\mathrm{seg}}+1.
\end{align}
Then, the extraction interval for vehicle $v$ is defined as
\begin{align}
    \mathcal{I}_v
    =
    \{T_0^{(v)},T_0^{(v)}+1,\ldots,T_0^{(v)}+T_{\mathrm{seg}}-1\}.
\end{align}
The set of non-depot sites included in the extraction intervals is defined as
\begin{align}
    \mathcal{N}_{\mathrm{sub}}
    =
    \left\{
        r_v^{(m)}(t)
        \mid
        v\in\mathcal{V}_{\mathrm{sub}},\
        t\in\mathcal{I}_v
    \right\}.
\end{align}
Because the current solution satisfies constraint A, the same non-depot site does not appear at different extraction steps.
Therefore,
\begin{align}
    |\mathcal{N}_{\mathrm{sub}}|=V'T_{\mathrm{seg}}.
\end{align}
The free variables of the subproblem are $x_{t,i}^{(v)}$ with $v\in\mathcal{V}_{\mathrm{sub}}$, $t\in\mathcal{I}_v$, and $i\in\mathcal{N}_{\mathrm{sub}}$.
All the other variables are fixed to the values of the current solution.
In LNS-VT, the subproblem is constructed using the set of selected vehicles $\mathcal{V}_{\mathrm{sub}}$, the extraction intervals $\{\mathcal{I}_v\}_{v\in\mathcal{V}_{\mathrm{sub}}}$, and the set of extracted non-depot sites $\mathcal{N}_{\mathrm{sub}}$.
Figure~\ref{fig:VRP_sub}(b) shows a schematic of the subproblem in LNS-VT.

The objective function of LNS-VT must include not only the travel distances within the extraction intervals but also the connection distances to the fixed sites immediately before and after the extraction intervals.
For vehicle $v$, we denote the fixed site immediately before the extraction interval by $a_v$ and the fixed site immediately after the extraction interval by $b_v$, which are given by
\begin{align}
    a_v = r_v^{(m)}(T_0^{(v)}-1),
    \qquad
    b_v = r_v^{(m)}(T_0^{(v)}+T_{\mathrm{seg}}).
\end{align}
The site $a_v$ or $b_v$ can be the depot depending on the position of the extraction interval.
The travel distances outside the extraction intervals are constants that do not depend on the optimization of the subproblem, and thus they can be omitted from the QUBO.
The objective function that depends on the free variables is given by
\begin{align}
    \label{eq: lnsvt_obj}
    \HVRPppobj(\boldsymbol{x})
    =
    \sum_{v\in\mathcal{V}_{\mathrm{sub}}}
    \Bigg[
    &
    \sum_{j\in\mathcal{N}_{\mathrm{sub}}}
    l_{a_v,j}x_{T_0^{(v)},j}^{(v)}
    \nonumber\\
    &+
    \sum_{t=T_0^{(v)}}^{T_0^{(v)}+T_{\mathrm{seg}}-2}
    \sum_{i\in\mathcal{N}_{\mathrm{sub}}}
    \sum_{j\in\mathcal{N}_{\mathrm{sub}}}
    l_{i,j}x_{t,i}^{(v)}x_{t+1,j}^{(v)}
    \nonumber\\
    &+
    \sum_{i\in\mathcal{N}_{\mathrm{sub}}}
    l_{i,b_v}x_{T_0^{(v)}+T_{\mathrm{seg}}-1,i}^{(v)}
    \Bigg].
\end{align}
The first term represents the travel distance from the fixed site immediately before the extraction interval to the site visited first in the extraction interval.
The second term represents the travel distances between consecutive steps within the extraction interval.
The third term represents the travel distance from the site visited last in the extraction interval to the fixed site immediately after the extraction interval.
When $T_{\mathrm{seg}}=1$, the sum in the second term is an empty sum and is regarded as zero.

In the subproblem of LNS-VT, we impose two constraints.
First, each non-depot site in $\mathcal{N}_{\mathrm{sub}}$ must be visited exactly once within the extraction intervals of the selected vehicles.
Second, each selected vehicle must visit exactly one non-depot site at each extracted step.
The QUBO energy function for the subproblem of LNS-VT is then given by
\begin{align}
    \label{eq: lnsvt_energy}
    \HVRPpp(\boldsymbol{x})
    =
    \HVRPppobj(\boldsymbol{x})
    &+
    \mu_\mathrm{A}
    \sum_{i\in\mathcal{N}_{\mathrm{sub}}}
    \left(
        \sum_{v\in\mathcal{V}_{\mathrm{sub}}}
        \sum_{t\in\mathcal{I}_v}
        x_{t,i}^{(v)}
        -1
    \right)^2
    \nonumber\\
    &+
    \mu_\mathrm{B}
    \sum_{v\in\mathcal{V}_{\mathrm{sub}}}
    \sum_{t\in\mathcal{I}_v}
    \left(
        \sum_{i\in\mathcal{N}_{\mathrm{sub}}}
        x_{t,i}^{(v)}
        -1
    \right)^2.
\end{align}
Equation~\eqref{eq: lnsvt_energy} does not include a penalty term corresponding to constraint C.
This is because the extraction intervals do not include steps at which a vehicle stays at the depot, and the free variables inside the intervals correspond only to non-depot sites.
The steps outside the intervals are fixed to the values of the current solution.
Therefore, the property that a vehicle does not leave the depot after returning to it is inherited from the current solution.
The depot visits at the first and last steps corresponding to constraint D also remain fixed.

This construction preserves feasibility in the following sense.
For any feasible solution of the LNS-VT subproblem, the first penalty term in Eq.~\eqref{eq: lnsvt_energy} is zero, and hence each non-depot site in $\mathcal{N}_{\mathrm{sub}}$ is visited exactly once within the extraction intervals of the selected vehicles.
The variables associated with the non-depot sites not included in $\mathcal{N}_{\mathrm{sub}}$ are all fixed to the values of the current solution.
Therefore, every non-depot site is visited exactly once after reinsertion, and constraint A is satisfied.
Next, the second penalty term in Eq.~\eqref{eq: lnsvt_energy} is zero for any feasible subproblem solution, and hence each selected vehicle visits exactly one non-depot site at each step within its extraction interval, while its variables at the other sites, including the depot, are fixed to zero at these steps.
Since the steps outside the extraction intervals are fixed to the current solution, constraint B remains satisfied after reinsertion.
Furthermore, the extraction intervals lie inside the consecutive non-depot part of each selected route.
Thus, reinsertion does not introduce a departure from the depot after a vehicle has returned to the depot.
Outside the intervals, the routes of the current solution are kept unchanged, and therefore constraint C is maintained.
Constraint D is also maintained by fixing the first and last steps.
Consequently, reinserting any feasible subproblem solution of LNS-VT into the current solution yields a feasible solution of the original problem.

Finally, we consider the number of variables in the subproblem of LNS-VT.
The number of selected vehicles is $V'$, the actual number of steps extracted from each selected vehicle is $T_{\mathrm{seg}}$, and the number of extracted non-depot sites is $|\mathcal{N}_{\mathrm{sub}}|=V'T_{\mathrm{seg}}$.
Because LNS-VT contains no free variables corresponding to the depot, the total number of free binary variables is
\begin{align}
    \label{eq: lnsvt_variable_number}
    n_{\mathrm{var}}^{\mathrm{VT}}
    =
    V'T_{\mathrm{seg}}|\mathcal{N}_{\mathrm{sub}}|
    =
    (V'T_{\mathrm{seg}})^2.
\end{align}
When no truncation of the prescribed segment length occurs, $T_{\mathrm{seg}}=T'$ and Eq.~\eqref{eq: lnsvt_variable_number} becomes
\begin{align}
    n_{\mathrm{var}}^{\mathrm{VT}}
    =
    (V'T')^2.
\end{align}
In LNS-V, the number of steps in the subproblem is fixed to the number of steps $T$ of the original problem.
In contrast, LNS-VT introduces the segment length $T'$ as an additional control parameter.
As a result, the subproblem size can be controlled more finely, and its dependence on the number of steps $T$ of the original problem is relaxed.
This is the main difference between LNS-VT and LNS-V.

Algorithm~\ref{alg:lns-vt} shows the subproblem generation procedure in LNS-VT.

\begin{algorithm}[t]
    \caption{Subproblem generation in LNS-VT}
    \label{alg:lns-vt}
    \begin{algorithmic}[1]
        \Require the number of selected vehicles $V'$, the prescribed segment length $T'$, current solution $\boldsymbol{x}^{(m)}$
        \Ensure set of selected vehicles $\mathcal{V}_{\mathrm{sub}}$, set of target sites $\mathcal{N}_{\mathrm{sub}}$, extraction intervals $\{\mathcal{I}_v\}_{v\in\mathcal{V}_{\mathrm{sub}}}$
        \For{$v\in\mathcal{V}$}
            \State Compute $T_{\mathrm{valid}}^{(v)}$
        \EndFor
        \State $\mathcal{V}_{+}^{(m)}\leftarrow\{v\in\mathcal{V}\mid T_{\mathrm{valid}}^{(v)}\geq 1\}$
        \State Select $V'$ vehicles uniformly at random without replacement from $\mathcal{V}_{+}^{(m)}$ and denote the set by $\mathcal{V}_{\mathrm{sub}}$
        \State $T_{\mathrm{seg}}\leftarrow \min\left(T',\min_{v\in\mathcal{V}_{\mathrm{sub}}}T_{\mathrm{valid}}^{(v)}\right)$
        \State $\mathcal{N}_{\mathrm{sub}}\leftarrow\emptyset$
        \For{$v\in\mathcal{V}_{\mathrm{sub}}$}
            \State Choose $T_0^{(v)}$ uniformly at random from $\{1,\ldots,T_{\mathrm{valid}}^{(v)}-T_{\mathrm{seg}}+1\}$
            \State $\mathcal{I}_v \leftarrow \{T_0^{(v)},T_0^{(v)}+1,\ldots,T_0^{(v)}+T_{\mathrm{seg}}-1\}$
            \State $\mathcal{N}^{(v)} \leftarrow \{r_v^{(m)}(t)\mid t\in\mathcal{I}_v\}$
            \State $\mathcal{N}_{\mathrm{sub}} \leftarrow \mathcal{N}_{\mathrm{sub}}\cup\mathcal{N}^{(v)}$
        \EndFor
        \State \Return $\mathcal{V}_{\mathrm{sub}}$, $\mathcal{N}_{\mathrm{sub}}$, $\{\mathcal{I}_v\}_{v\in\mathcal{V}_{\mathrm{sub}}}$
    \end{algorithmic}
\end{algorithm}

\section{Results}
\label{sec: result}

This section reports the numerical results for the VRP.
In Sec.~\ref{subsec: setting}, we describe the settings of the numerical experiments.
In Sec.~\ref{subsec: result_VRP}, we compare LNS-VT with LNS-V and the naive method, and examine how the final objective value depends on the subproblem size.

\subsection{Numerical Experiment Settings}
\label{subsec: setting}

\begin{table}[b]
  \centering
  \setlength{\abovecaptionskip}{4pt}
  \setlength{\belowcaptionskip}{6pt}
  \caption{Conditions of the target VRP and parameter settings used for LNS-VT.}
  \label{tab: VRPsettings}
  \begin{tabular}{cc}
      \toprule
      Parameter & Value \\
      \hline
      Number of depots & 1 \\
      $N$: Number of non-depot sites & 300 \\
      $V$: Number of vehicles & 5 \\
      $T$: Number of steps & 62 \\
      \hline
      \multirow{4}{*}{\shortstack[c]{Number of selected vehicles $V'$ and \\ prescribed segment length $T'$\\
       in LNS-VT}} 
      & $V'=2$, $T'=\{10, 20, 30, 40, 50\}$\\
      & $V'=3$, $T'=\{10, 20, 30, 40\}$\\
      & $V'=4$, $T'=\{10, 15, 20, 25, 30\}$\\
      & $V'=5$, $T'=\{12, 16, 20, 24\}$\\
      \bottomrule
  \end{tabular}
\end{table}

Table~\ref{tab: VRPsettings} shows the VRP instance and the parameter settings used for LNS-VT.
The non-depot sites were distributed uniformly at random in a unit square, and the depot was placed at the center of the square.
We set the number of steps to
\begin{align}
    T=\left\lceil \frac{N}{V}\right\rceil +2 = 62,
\end{align}
where the additional two steps correspond to the departure from and return to the depot.
In this setting, $N=V(T-2)$ holds.
Therefore, in any feasible solution, each vehicle visits exactly $T-2=60$ non-depot sites, and $T_{\mathrm{valid}}^{(v)}=60$ holds for every vehicle.
Because all prescribed values of $T'$ in Table~\ref{tab: VRPsettings} satisfy $T'\leq 60$, no truncation of the prescribed segment length occurs in LNS-VT, and $T_{\mathrm{seg}}=T'$ holds in all runs.

For each parameter setting, we performed 10 runs and evaluated the solution quality using the average objective value.
The error bars in the figures represent the standard deviations over these 10 runs.
For the naive method, LNS-V, and LNS-VT, every penalty coefficient in Eqs.~\eqref{eq: VRPenergy}, \eqref{eq: lns_v_energy}, and \eqref{eq: lnsvt_energy} was set to $l_{\max}=\max_{i,j\in\mathcal{N}_0}l_{i,j}$, the maximum inter-site distance of the original instance~\cite{lucas2014ising}.
In particular, the same value $l_{\max}$ was used for all subproblem QUBOs without recomputation.
After each run, the feasibility of the obtained solution was checked.

For subproblem generation, the selected vehicles were sampled uniformly at random without replacement.
For LNS-VT, the extraction start step $T_0^{(v)}$ was also sampled uniformly at random from the admissible range for each selected vehicle.
The target sites in the subproblem were then determined by the current solution and the selected extraction intervals.
For LNS-V, we used $V'=2$ as the baseline setting, following the parameter setting that gave the best solution quality in the previous study~\cite{yamashita2026Geometric}.

To remove the effect of different initial solutions, all LNS-V and LNS-VT runs were started from the same feasible initial solution.
This initial solution was a single solution obtained by the naive method.
Thus, the results below evaluate how each LNS setting improves this common initial solution.

\begin{table}[t]
  \centering
  \caption{Calculation conditions of LNS-VT, LNS-V, and the naive method for the VRP.}
  \label{tab: AEmethod_VRP}
  \begin{tabular}{p{2cm} p{2.5cm} p{2.5cm}}
      \toprule
      \parbox[c]{2cm}{\centering Algorithm}
      & \parbox[c]{2.5cm}{\centering Number of QUBOs\\ solved}
      & \parbox[c]{2.5cm}{\centering Computation time \\ per QUBO} \\
      \hline
      \parbox[c]{2cm}{\centering LNS-VT}
      & \parbox[c]{2.5cm}{\centering 100}
      & \parbox[c]{2.5cm}{\centering 10 s} \\
      \parbox[c]{2cm}{\centering LNS-V}
      & \parbox[c]{2.5cm}{\centering 100}
      & \parbox[c]{2.5cm}{\centering 10 s} \\
      \parbox[c]{2cm}{\centering Naive method}
      & \parbox[c]{2.5cm}{\centering 1}
      & \parbox[c]{2.5cm}{\centering 300 s} \\
      \bottomrule
  \end{tabular}
\end{table}

Table~\ref{tab: AEmethod_VRP} shows the calculation conditions for LNS-VT, LNS-V, and the naive method.
The naive method directly solves the full QUBO formulation of the original VRP without LNS-based problem reduction.
Because this optimization is performed only once, the number of QUBOs solved is 1.
The computation time for the naive method was set to be long enough that no further improvement in solution quality was observed with longer computation times.
In contrast, LNS-VT and LNS-V solve one subproblem QUBO at each iteration.

The total computation time on the Ising machine is therefore 1000 s for each LNS run and 300 s for each naive-method run.
Consequently, the comparison with the naive method should be interpreted as a comparison of solution quality under the stated computational budgets, not as a wall-clock-time comparison.
On the other hand, LNS-VT and LNS-V are compared under the same number of iterations and the same computation time per subproblem QUBO.
In this study, Fixstars Amplify AE~\cite{FAAE} was used as the Ising machine.

\subsection{Results for the VRP}
\label{subsec: result_VRP}

\begin{figure}
  \centering
  \subfloat[]{\includegraphics[width=0.85\linewidth]{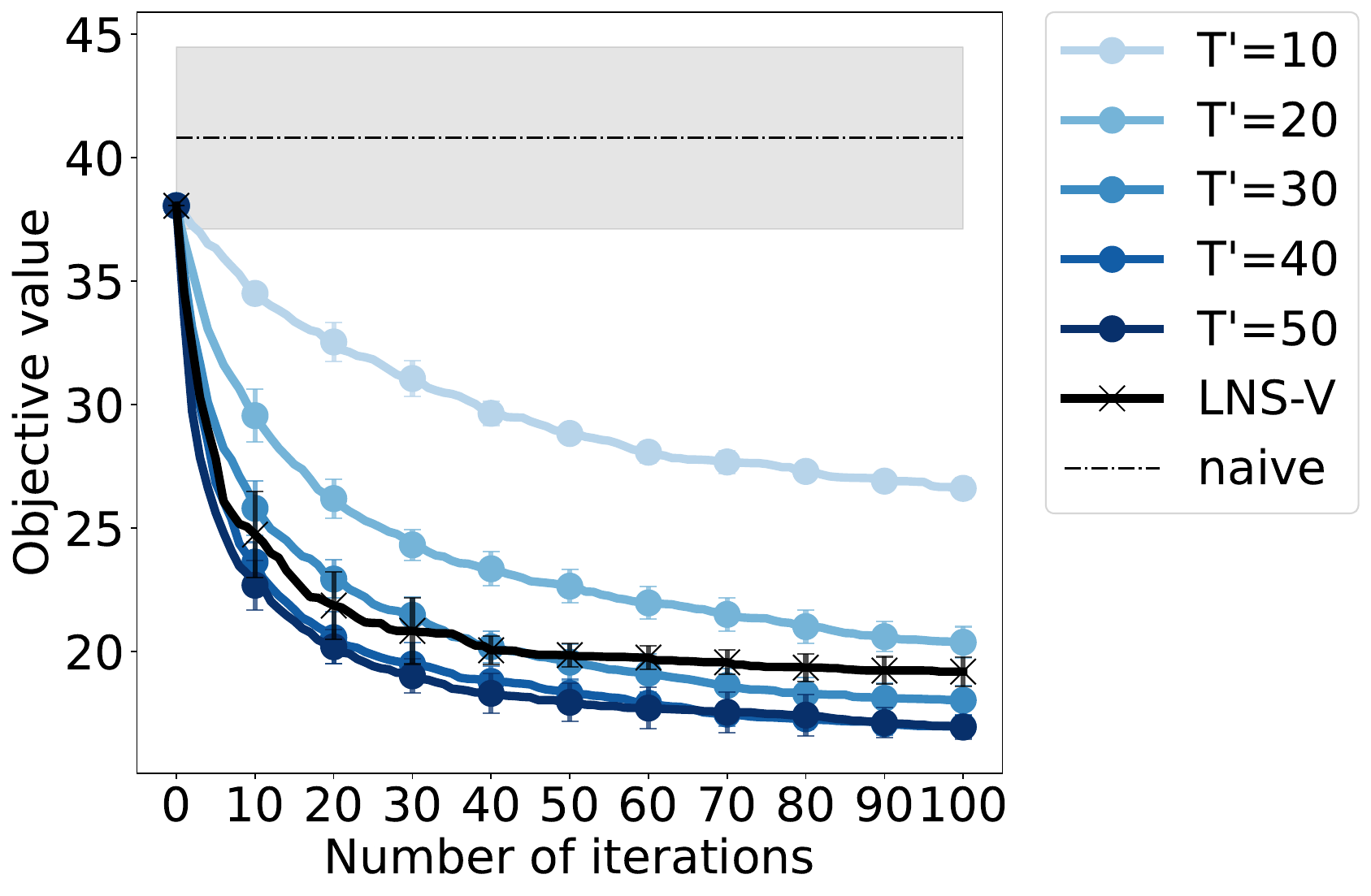}\label{fig:sub_a}}\\
  \subfloat[]{\includegraphics[width=0.85\linewidth]{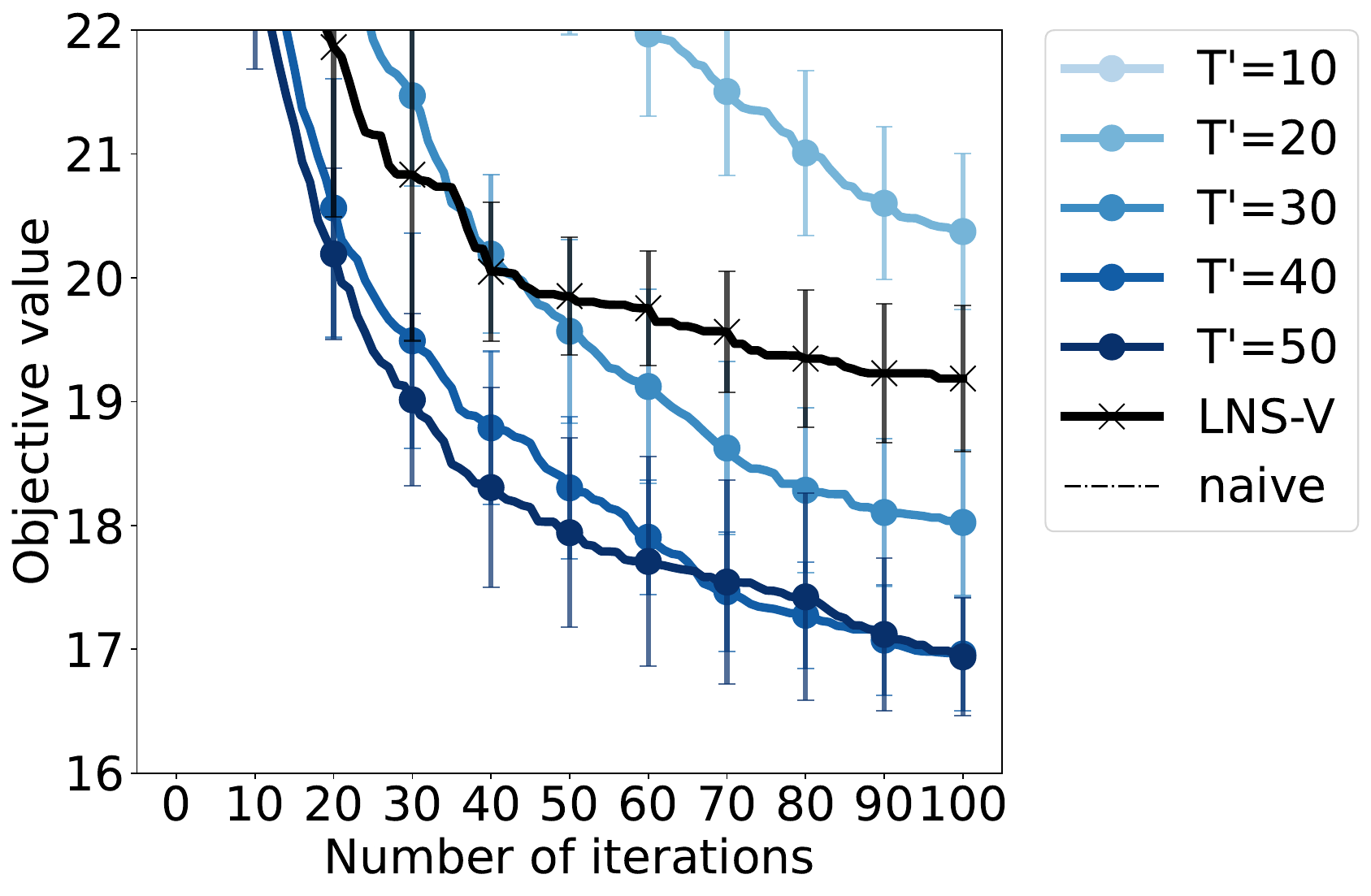}\label{fig:sub_b}}
  \caption{Update process of the objective value in LNS for the VRP with 300 non-depot sites and 5 vehicles. The results of LNS-VT are indicated by ``$T'=$''. Panel (b) is an enlarged view of panel (a) along the vertical axis. The points represent the average values over 10 runs, and the error bars represent the standard deviations.}
  \label{fig: VRP300_5_2}
\end{figure}

Figure~\ref{fig: VRP300_5_2} shows the average objective values obtained by LNS-VT, LNS-V, and the naive method for the VRP with 300 non-depot sites and 5 vehicles.
For LNS-VT and LNS-V, the horizontal axis represents the number of LNS iterations.
For the naive method, the dashed line represents the average objective value obtained by directly solving the full QUBO.
The results of LNS-VT are indicated by ``$T'=$'' in the legend.

The common initial solution used for LNS had a lower objective value than the average value obtained by the naive method.
Therefore, the dashed line for the naive method should be interpreted as the average performance of direct full-QUBO solving, not as the objective value of the initial solution used in LNS.
Both LNS-V and LNS-VT decreased the objective value from the common initial solution in the tested instance.
In the reported runs, the empirical feasible-solution rate was 100\% for the naive method, LNS-V, and LNS-VT.

Next, we compare LNS-V and LNS-VT for $V'=2$.
For LNS-VT with $T'=10$ and $T'=20$, the objective value decreased more slowly than for LNS-V, and the final average objective values after 100 iterations were not lower than that of LNS-V.
In contrast, LNS-VT with $T'\geq 30$ yielded lower final average objective values than LNS-V.
Among the tested $V'=2$ settings, LNS-VT with $T'=40$ gave the lowest final average objective value.
At 100 iterations, this setting reduced the average objective value by approximately 10\% relative to LNS-V.
Moreover, after approximately 30 iterations, the $T'=40$ curve reached an objective value comparable to the final average objective value of LNS-V.
Thus, under the same computation time per QUBO, the best tested LNS-VT setting improved the final solution quality and reached the final quality of LNS-V with fewer iterations.
This comparison does not by itself establish an end-to-end wall-clock-time advantage.

\begin{figure}[t]
  \centering
  \includegraphics[width=0.95\linewidth]{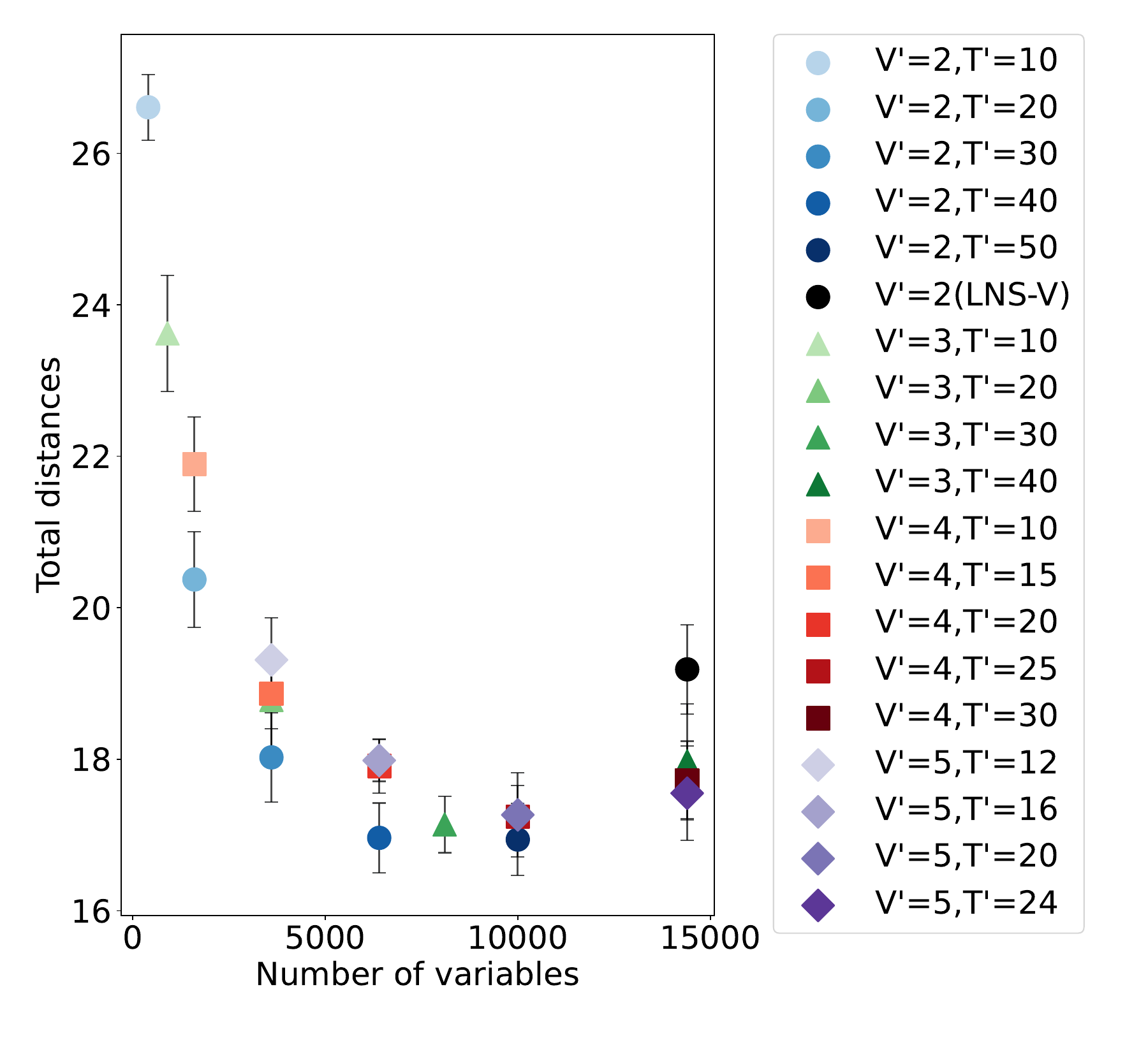}
  \caption{Relation between the number of non-depot site-assignment variables included in the subproblem and the objective value after 100 iterations of LNS for the VRP with 300 non-depot sites and 5 vehicles. The points represent the average values over 10 runs, and the error bars represent the standard deviations.}
  \label{fig: sub_variables}
\end{figure}

Figure~\ref{fig: sub_variables} shows the relation between the number of variables included in the subproblem and the objective value after 100 iterations.
Here, the number of variables denotes the number of non-depot site-assignment variables introduced in Sec.~\ref{sec: method}.
For LNS-VT, this number is $(V'T')^2$ in the present experiment because $T_{\mathrm{seg}}=T'$ and no depot variables are included as free variables.
For LNS-V, the value is computed from Eq.~\eqref{eq: lns_v_variable_number}, which applies here because each vehicle visits exactly $T-2$ non-depot sites in the present setting.

Under the problem instance and Ising-machine settings used in this study, the lowest final average objective values were obtained when the number of variables in the subproblem was approximately 6000 to 10000.
When the number of variables deviated from this range, the final average objective value tended to increase.
This increase was particularly clear for settings with fewer variables.

We next compare this range with the subproblem size available to LNS-V.
Under the restriction $V'\geq 2$ adopted in this study, the minimum number of selected vehicles in LNS-V is $V'=2$.
Using Eq.~\eqref{eq: lns_v_variable_number} with $T=62$, the corresponding number of non-depot site-assignment variables is
\begin{align}
    n_{\mathrm{var}}^{\mathrm{V}}
    =
    2^2(62-2)^2
    =
    14400.
\end{align}
Thus, under the present setting and the restriction $V'\geq 2$, LNS-V cannot realize subproblems with approximately 6000 to 10000 non-depot site-assignment variables.
This observation provides a possible explanation for why the best tested LNS-VT setting outperformed LNS-V in this experiment.
It also supports the motivation for introducing $T'$ as an additional parameter for controlling the subproblem size.
Because the LNS-V subproblem size scales as $(V')^2(T-2)^2$, this limitation becomes more severe as the number of steps $T$ increases.

\section{Discussion}
\label{sec: discussion}

In this section, we discuss the VRP results presented in Sec.~\ref{sec: result}.
The discussion is limited to the problem instance, parameter settings, and Ising-machine settings used in the numerical experiments.
In Sec.~\ref{subsec: discussion_variable}, we examine how the appropriate subproblem size changes during the LNS iterations.
In Sec.~\ref{subsec: discussion_v}, we discuss why different combinations of the number of selected vehicles $V'$ and the prescribed segment length $T'$ can lead to different solution qualities even when the number of variables in the subproblem is the same.

\subsection{Relation between the Number of Variables and Solution Quality}
\label{subsec: discussion_variable}

The results in Fig.~\ref{fig: VRP300_5_2} suggest that a subproblem setting that is effective in the early stage of LNS is not necessarily the most effective in the later stage.
For example, when $V'=2$, the average objective-value curves for LNS-VT with $T'=30$ and for LNS-V cross at approximately 50 iterations.
Thus, the relative performance of these two settings changes during the update process.

\begin{figure}[t]
  \centering
  \includegraphics[width=\linewidth]{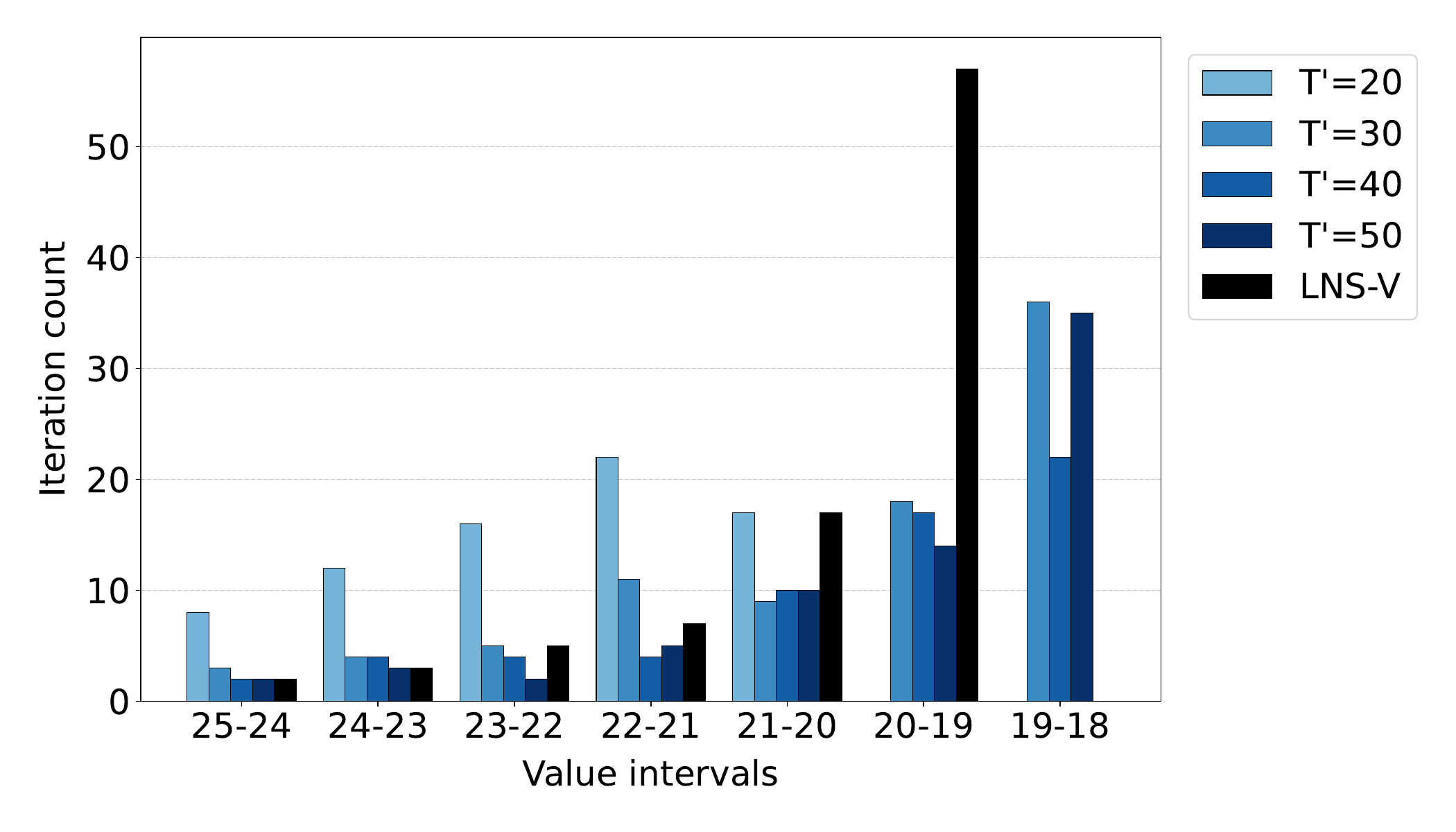}
  \caption{Number of iterations required for the average objective value to pass through each objective-value interval for LNS-V and LNS-VT with $V'=2$. The horizontal axis represents objective-value intervals, and the vertical axis represents the number of iterations required to pass through each interval. A smaller value indicates that the average objective value decreased more quickly in that interval.}
  \label{fig: iteration_count}
\end{figure}

To examine this behavior more closely, we counted the number of iterations required for the average objective value of each setting to pass through specified objective-value intervals.
The results are shown in Fig.~\ref{fig: iteration_count}.
We used unit-width intervals from 25--24 to 19--18.
For each interval, the count represents the number of iterations during which the 10-run average objective value stayed within the half-open interval $[\text{low}, \text{high})$, bounded by the lower and upper endpoints of that interval.
This count practically corresponds to the number of iterations required to pass through the interval.
If the average objective value did not reach the upper endpoint of a given interval within 100 iterations, the setting was excluded from the plot for that interval.
Conversely, if the average objective value reached the upper endpoint but failed to drop below the lower endpoint within 100 iterations, the count includes all iterations from reaching the upper endpoint up to the 100 iteration.
A smaller count means that the average objective value decreased more quickly in that objective-value range.
Therefore, this count is used here as an empirical indicator of how effective each subproblem setting is at a given stage of the search.

Figure~\ref{fig: iteration_count} shows that, in the intervals with larger objective values, such as 25--24 and 24--23, settings with a larger number of variables tend to require fewer iterations.
In contrast, as the objective value decreases, the setting with the smallest iteration count shifts toward a smaller number of variables.
This observation suggests that the appropriate subproblem size changes with the quality of the current solution.
In the early stage, a larger subproblem can contain solutions with larger improvements, and this can accelerate the decrease in the objective value.
In the later stage, however, the current solution has already been improved, and the remaining improvements become smaller.
In this regime, the quality of the solution returned by the Ising machine for each subproblem becomes more important.
A smaller subproblem may then be advantageous because it can be solved more accurately within the same computation time.
This interpretation is consistent with the observed shift in Fig.~\ref{fig: iteration_count}, although the present data do not determine an automatic rule for choosing the subproblem size.

\subsection{Effect of the Number of Selected Vehicles}
\label{subsec: discussion_v}

Figure~\ref{fig: sub_variables} shows that, among settings with the same number of non-depot site-assignment variables, the final average objective value also depends on the combination of $V'$ and $T'$.
As described in Sec.~\ref{subsec: setting}, no truncation of the prescribed segment length occurs in the present VRP experiment, and the number of variables in LNS-VT is $(V'T')^2$.
For a fixed value of $(V'T')^2$, a smaller $V'$ corresponds to a larger $T'$.
In the tested settings, smaller $V'$ tended to give lower final average objective values when the number of variables was small, whereas larger $V'$ tended to give lower final average objective values when the number of variables was large.
We discuss this trend from two viewpoints, namely the direct effect of the number of vehicles on the QUBO solution quality and the composition of the route segments included in the subproblem.

\begin{figure}
  \centering
  \includegraphics[width=0.85\linewidth]{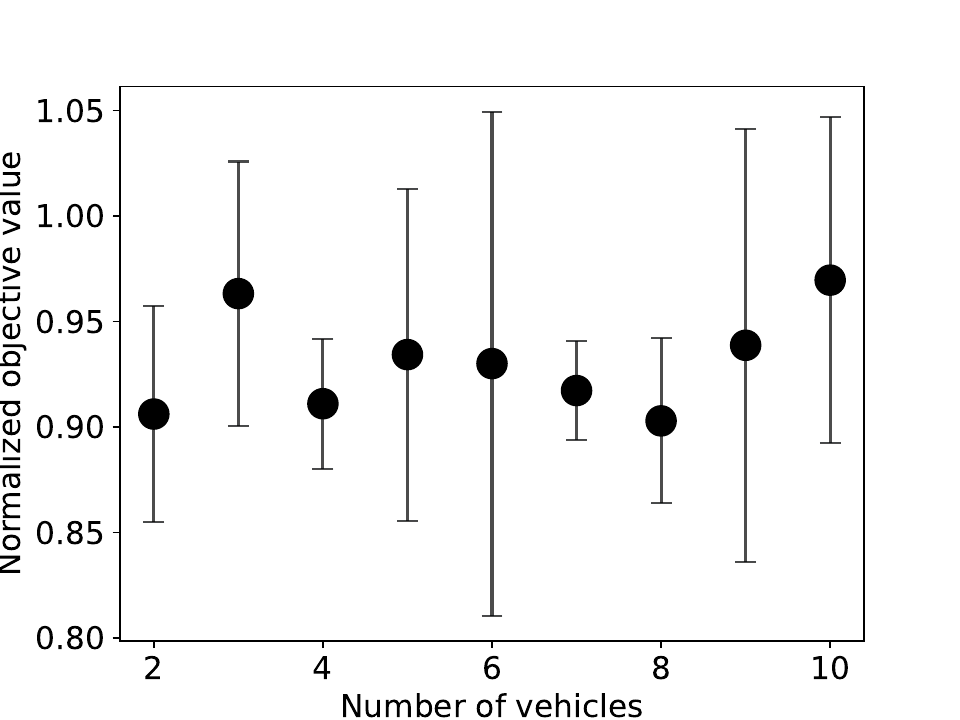}
  \caption{Relation between the total number of vehicles in the VRP and the objective value normalized by the greedy solution. The number of non-depot sites was fixed at 120, and the number of vehicles was varied from 2 to 10. The full QUBO was solved directly 10 times for each setting. The points represent the average values, and the error bars represent the standard deviations.}
  \label{fig: v_dependence}
\end{figure}

First, we examine whether the number of vehicles itself strongly affects the solution quality returned by the Ising machine.
For this purpose, we generated VRP instances with 120 non-depot sites, while the total number of vehicles was varied from 2 to 10. 
The non-depot sites were distributed in the same manner as in Sec.~\ref{subsec: setting}, and the same site configuration was used across all vehicle numbers. 
For each vehicle number, the number of steps was set to $T=\lceil N/V\rceil+2$. 
The penalty coefficients were set in the same manner as in Sec.~\ref{subsec: setting}. 
For each number of vehicles, the full QUBO was solved directly without LNS, the computation time was set to 10~s, and 10 runs were performed.
Figure~\ref{fig: v_dependence} shows the resulting objective values normalized by the value of the greedy solution.
Here, we constructed the greedy solution by repeatedly moving each vehicle from its current location to the nearest unvisited site. 
Specifically, the algorithm proceeds to construct the route for the next vehicle only after the current vehicle has visited $T$ sites and returned to the depot. 
Furthermore, if no unvisited sites remain, the vehicle stays at the depot.
No clear monotonic dependence on the total number of vehicles is visible in Fig.~\ref{fig: v_dependence}, and the error bars overlap for many of the tested vehicle numbers.
Thus, within this auxiliary experiment, the number of vehicles itself does not appear to be the main factor explaining the dependence on $V'$ observed in Fig.~\ref{fig: sub_variables}.

\begin{figure}
  \centering
  \subfloat[]{\includegraphics[width=0.90\linewidth]{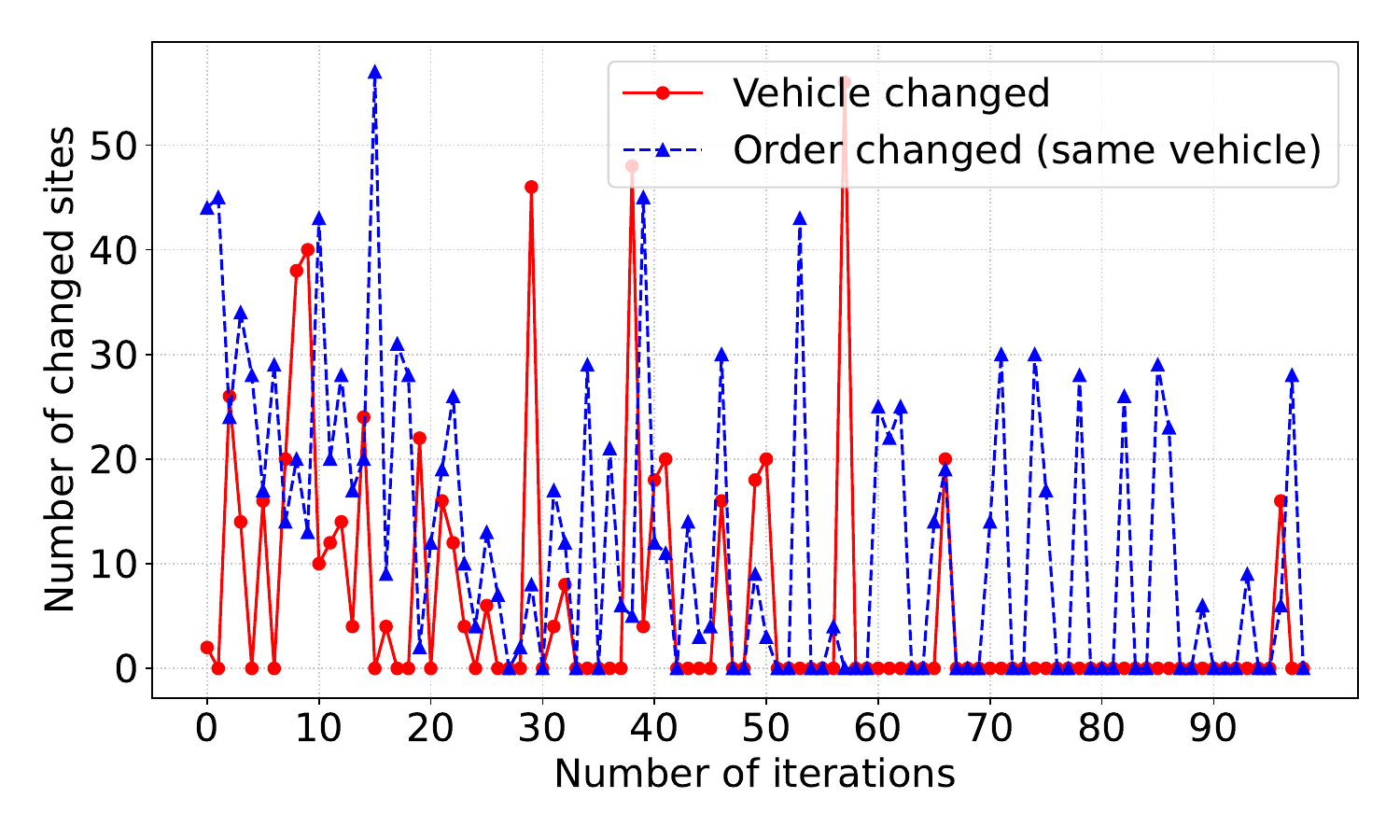}}\label{fig: node_a}\\
  \subfloat[]{\includegraphics[width=0.90\linewidth]{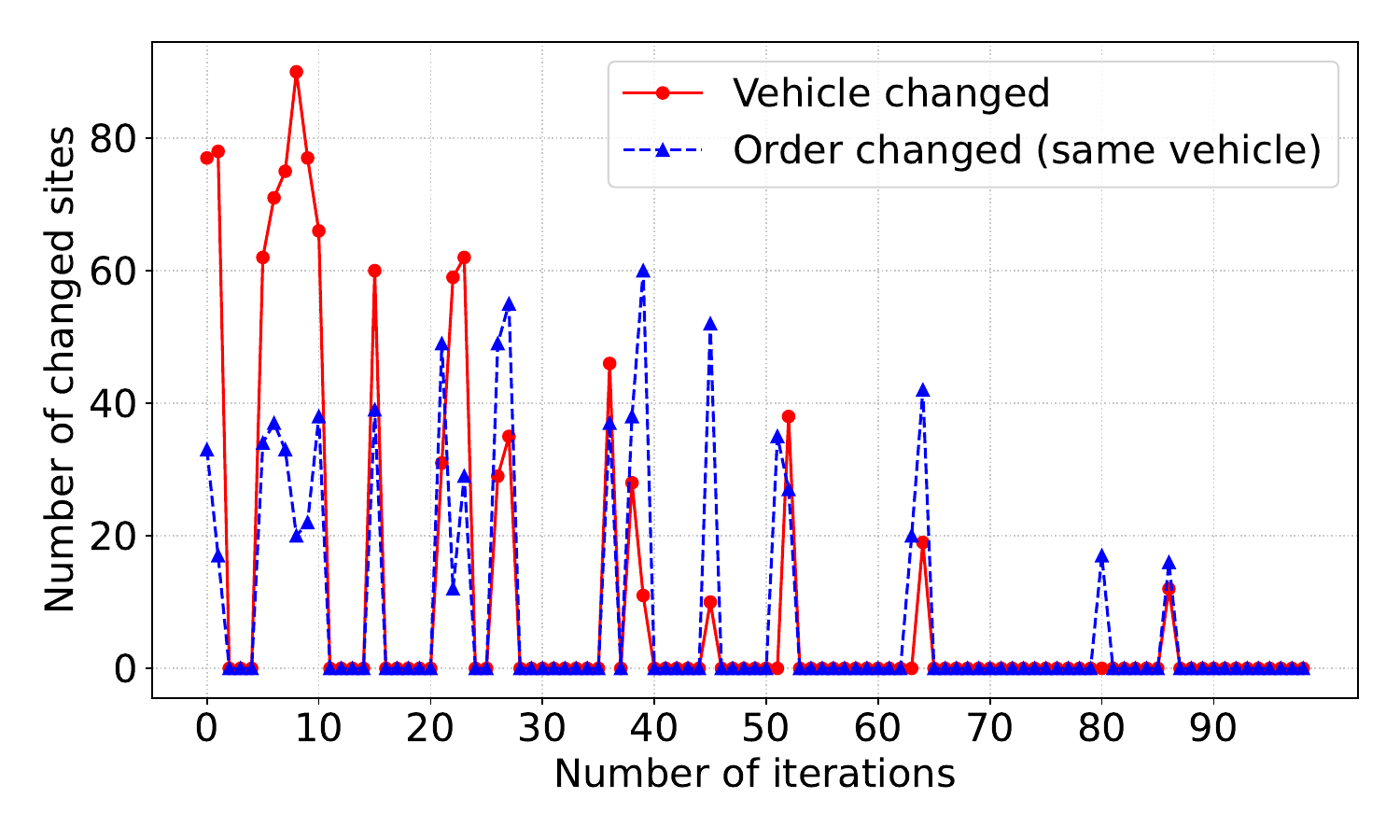}}\label{fig: node_b}
  \caption{Relation between the number of iterations in LNS-VT and the number of non-depot sites whose visiting vehicle or visiting order changed. Red indicates the number of sites whose visiting vehicle changed, and blue indicates the number of sites whose visiting vehicle remained the same but whose visiting order changed. (a) Results for $V'=2$ and $T'=30$. (b) Results for $V'=5$ and $T'=24$.}
  \label{fig: node_change}
\end{figure}

Next, we examine how the route segments included in the LNS-VT subproblem affect the update.
For each iteration, we counted the number of non-depot sites whose visiting vehicle or visiting order changed between the current solution and the updated solution.
We classified these changed sites into sites whose visiting vehicle changed and sites whose visiting vehicle remained the same but whose visiting order changed.
The results are shown in Fig.~\ref{fig: node_change}.
Figure~\ref{fig: node_change}(a) shows a representative small-subproblem setting, $V'=2$ and $T'=30$, for which the number of variables is 3600.
Figure~\ref{fig: node_change}(b) shows a representative large-subproblem setting, $V'=5$ and $T'=24$, for which the number of variables is 14400.

In Fig.~\ref{fig: node_change}(a), many updates involve changes in the visiting order within the same vehicle.
In Fig.~\ref{fig: node_change}(b), especially in the early iterations, many updates involve changes in the visiting vehicle.
This qualitative tendency was also observed in the other tested settings.
These results suggest that the effect of the subproblem setting is not determined solely by the total number of variables.
For a fixed value of $(V'T')^2$, a smaller $V'$ gives a longer contiguous route segment for each selected vehicle.
This can be beneficial when the main improvement is obtained by reordering sites within a vehicle route.
In contrast, with a larger $V'$, segments from more vehicles are included in the subproblem.
This can be beneficial when the main improvement is obtained by reassigning sites among different vehicles.

Therefore, the dependence on $V'$ observed in Fig.~\ref{fig: sub_variables} can be interpreted as a consequence of two effects.
The parameter $T'$ controls the length of the route segment optimized within each selected vehicle, whereas $V'$ controls how many vehicles can exchange sites.
Which of these effects is more important appears to depend on the subproblem size and on the stage of the LNS iterations.
This observation supports the use of both $V'$ and $T'$ as control parameters in LNS-VT.
At the same time, the present results are based on a limited set of VRP instances and parameter settings.
A systematic adaptive strategy for selecting $V'$ and $T'$ remains a topic for future work.

\section{Extension to the Quadratic Multiple Knapsack Problem}
\label{sec: QMKP}

In this section, we examine whether the design principle used in LNS-VT, namely controlling the subproblem size while preserving feasibility, can also be applied to a constrained combinatorial optimization problem other than the VRP.
As an example, we consider the QMKP.
In Sec.~\ref{subsec: QMKP}, we summarize the QMKP and its QUBO formulation.
In Sec.~\ref{subsec: QMKP_proposed}, we describe the subproblem generation method for the QMKP.
In Sec.~\ref{subsec: QMKP_setting}, we describe the numerical settings.
In Sec.~\ref{subsec: QMKP_result}, we present the numerical results.

\subsection{Quadratic Multiple Knapsack Problem}
\label{subsec: QMKP}

In the QMKP, $N$ items and $K$ knapsacks are given.
We denote the set of items by $\mathcal{N}=\{0,1,\ldots,N-1\}$ and the set of knapsacks by $\mathcal{K}=\{0,1,\ldots,K-1\}$.
Each item $i\in\mathcal{N}$ has a weight $w_i$, and each knapsack $k\in\mathcal{K}$ has a capacity $C_k$.
The value obtained by placing items in the same knapsack is specified by $p_{i,j}$.
Here, $p_{i,i}$ represents the value of item $i$, and $p_{i,j}$ for $i\neq j$ represents the correlated value of items $i$ and $j$.
The values $p_{i,j}$ are given for $i\leq j$, and we set $p_{i,j}=p_{j,i}$ for $i>j$.

The objective of the QMKP is to maximize the total value of the items placed in the knapsacks, under the constraint that the total weight of the items in each knapsack does not exceed its capacity.
Each item may be placed in at most one knapsack, and an item may also remain outside all knapsacks.

To formulate the QMKP as a QUBO, we introduce the following binary variables $x_{k,i}$.
\begin{equation}
    \label{eq: QMKP_variable}
    x_{k,i} =
    \begin{cases}
        1, & \text{if item $i$ is placed in knapsack $k$},\\
        0, & \text{otherwise}.
    \end{cases}
\end{equation}
Here, $k\in\mathcal{K}$ is the knapsack index, and $i\in\mathcal{N}$ is the item index.
The number of decision variables is $NK$.

In terms of the binary variables, the total profit is
\begin{align}
    \label{eq: QMKP_profit}
    P_{\mathrm{QMKP}}(\boldsymbol{x})
    =
    \sum_{k\in\mathcal{K}}
    \sum_{i\in\mathcal{N}}
    \sum_{\substack{j\in\mathcal{N}\\ j\ge i}}
    p_{i,j}x_{k,i}x_{k,j}.
\end{align}
Because the QMKP is a maximization problem, we minimize the negative profit when using an Ising machine.
Thus, the QUBO objective term is
\begin{align}
    \label{eq: QMKPobjective}
    \HQMKPobj(\boldsymbol{x})
    =
    -
    \sum_{k\in\mathcal{K}}
    \sum_{i\in\mathcal{N}}
    \sum_{\substack{j\in\mathcal{N}\\ j\ge i}}
    p_{i,j}x_{k,i}x_{k,j}.
\end{align}

The QMKP has the following two constraints.

\begin{description}
    \item[Constraint E]
    The total weight of the items placed in each knapsack must not exceed its capacity.
    \begin{align}
        \label{eq: QMKPconstE}
        \sum_{i\in\mathcal{N}}w_i x_{k,i}\leq C_k,
        \quad \forall k\in\mathcal{K}.
    \end{align}

    \item[Constraint F]
    Each item is placed in at most one knapsack.
    \begin{align}
        \label{eq: QMKPconstF}
        \sum_{k\in\mathcal{K}}x_{k,i}\leq 1,
        \quad \forall i\in\mathcal{N}.
    \end{align}
\end{description}

We assume that the weights and capacities are nonnegative integers, as in the benchmark instances used below.
Constraint E is an inequality constraint.
To express it in a QUBO form, we introduce auxiliary binary variables $y_{k,d}$.
Using log encoding, the capacity constraint for knapsack $k$ can be represented by
\begin{align}
    \sum_{d=0}^{D_k-1}2^d y_{k,d}
    -
    \left(2^{D_k}-1-C_k\right)
    -
    \sum_{i\in\mathcal{N}}w_i x_{k,i}
    =
    0,
\end{align}
where
\begin{align}
    D_k=\left\lceil \log_2(C_k+1)\right\rceil.
\end{align}
If the total weight does not exceed $C_k$, the auxiliary variables can be chosen so that the left-hand side of this equality is zero.
If the total weight exceeds $C_k$, no such choice exists.

Constraint F can be penalized by using
\begin{align}
    \left(\sum_{k\in\mathcal{K}}x_{k,i}\right)
    \left(\sum_{k\in\mathcal{K}}x_{k,i}-1\right),
\end{align}
which is zero when item $i$ is assigned to zero or one knapsack and positive when it is assigned to two or more knapsacks.

Combining the objective term with the penalty terms gives the QUBO energy function
\begin{align}
    \label{eq: QMKPenergy}
    \HQMKP(\boldsymbol{x},\boldsymbol{y})
    &=
    \HQMKPobj(\boldsymbol{x})
    \nonumber\\
    &+
    \mu_\mathrm{E}
    \sum_{k\in\mathcal{K}}
    \Bigg[
        \sum_{d=0}^{D_k-1}2^d y_{k,d}
        -
        \left(2^{D_k}-1-C_k\right)\nonumber\\
        &\qquad\qquad\qquad\qquad\quad-
        \sum_{i\in\mathcal{N}}w_i x_{k,i}
    \Bigg]^2
    \nonumber\\
    &+
    \mu_\mathrm{F}
    \sum_{i\in\mathcal{N}}
    \left(
        \sum_{k\in\mathcal{K}}x_{k,i}
    \right)
    \left(
        \sum_{k\in\mathcal{K}}x_{k,i}-1
    \right).
\end{align}
Here, $\mu_\mathrm{E}$ and $\mu_\mathrm{F}$ are positive penalty coefficients for constraints E and F, respectively.

\subsection{Subproblem Generation Method}
\label{subsec: QMKP_proposed}

We next describe a subproblem generation method for applying Ising-machine-assisted LNS to the QMKP.
The purpose of this construction is not to use LNS-VT itself, which is defined for the VRP, but to apply the same design principle to a problem with a different constraint structure.
In the following, we refer to Ising-machine-assisted LNS with this subproblem generation method as QMKP-LNS.

In the full QMKP formulation, the number of binary variables is
\begin{align}
    \label{eq: QMKP_variable_number_full}
    NK+\sum_{k\in\mathcal{K}}D_k.
\end{align}
The first term comes from the decision variables $x_{k,i}$, and the second term comes from the auxiliary variables for the capacity constraints.
Because $D_k$ depends logarithmically on $C_k$, the dominant contribution in the settings considered here is the decision-variable term $NK$.
Thus, the subproblem size can be controlled mainly by selecting the number of knapsacks and the number of items included in the subproblem.

Let $\boldsymbol{x}^{(m)}$ be a feasible current solution.
We first select $K'$ knapsacks uniformly at random without replacement from $\mathcal{K}$ and denote the selected knapsack set by $\mathcal{K}_{\mathrm{sub}}$.
We then define the candidate item set
\begin{align}
    \mathcal{N}_{\mathrm{cand}}^{(m)}
    =
    \left\{
        i\in\mathcal{N}
        \mid
        \sum_{k\in\mathcal{K}_{\mathrm{sub}}}x_{k,i}^{(m)}=1
        \ \mathrm{or}\
        \sum_{k\in\mathcal{K}}x_{k,i}^{(m)}=0
    \right\}.
\end{align}
This set consists of items currently placed in the selected knapsacks and items currently placed in no knapsack.
Items placed in knapsacks not selected for the subproblem are excluded from $\mathcal{N}_{\mathrm{cand}}^{(m)}$ and remain fixed.

From $\mathcal{N}_{\mathrm{cand}}^{(m)}$, we select
\begin{align}
    N_{\mathrm{sel}}=\min\left(N',|\mathcal{N}_{\mathrm{cand}}^{(m)}|\right)
\end{align}
items uniformly at random without replacement.
The selected item set is denoted by $\mathcal{N}_{\mathrm{sub}}$.
When $|\mathcal{N}_{\mathrm{cand}}^{(m)}|\geq N'$, we have $N_{\mathrm{sel}}=N'$.
The free decision variables of the subproblem are $x_{k,i}$ with $k\in\mathcal{K}_{\mathrm{sub}}$ and $i\in\mathcal{N}_{\mathrm{sub}}$.
All other decision variables are fixed to the values in the current solution.

\begin{figure}
  \centering
  \includegraphics[width=0.85\linewidth]{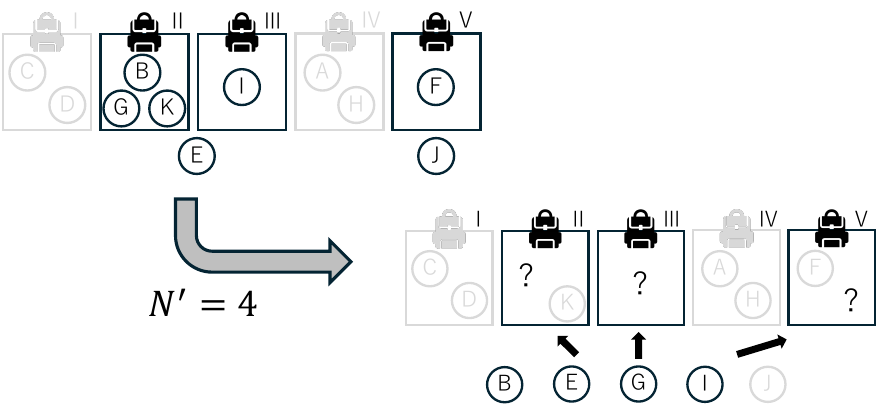}
  \caption{Schematic of the subproblem generation method for the QMKP. In the example with $K'=3$ and $N'=4$, knapsacks II, III, and V and items B, E, G, and I are included in the subproblem. The other items are fixed to their assignments in the current solution.}
  \label{fig: QMKP_subproblem}
\end{figure}

This construction allows the selected items to be reassigned among the selected knapsacks or to be removed from all selected knapsacks.
All items outside $\mathcal{N}_{\mathrm{sub}}$ are fixed to their current assignments.
This includes the items placed in unselected knapsacks, the items in the selected knapsacks that are not chosen for the subproblem, and the unassigned items that are not chosen.
A schematic of this construction is shown in Fig.~\ref{fig: QMKP_subproblem}.

To define the subproblem objective, we must include the interaction between free items and fixed items placed in the same selected knapsack.
For $k\in\mathcal{K}_{\mathrm{sub}}$, let
\begin{align}
    \mathcal{N}_{\mathrm{fix}}^{(m)}(k)
    =
    \left\{
        j\in\mathcal{N}\setminus\mathcal{N}_{\mathrm{sub}}
        \mid
        x_{k,j}^{(m)}=1
    \right\}
\end{align}
be the set of fixed items currently placed in knapsack $k$.
For $i\in\mathcal{N}_{\mathrm{sub}}$, we define
\begin{align}
    \label{eq: QMKP_u}
    u_i^{(k)}
    =
    \sum_{j\in\mathcal{N}_{\mathrm{fix}}^{(m)}(k)}p_{i,j}.
\end{align}
This quantity represents the profit contribution generated when item $i$ is placed in knapsack $k$ together with the fixed items already placed in that knapsack.

The part of the negative profit that depends on the free variables is
\begin{align}
    \label{eq: QMKP_sub_obj}
    {\mathcal{H}'}^{\mathrm{QMKP}}_{\mathrm{obj}}(\boldsymbol{x})
    =
    -
    \sum_{k\in\mathcal{K}_{\mathrm{sub}}}
    \left[
        \sum_{i\in\mathcal{N}_{\mathrm{sub}}}
        \sum_{\substack{j\in\mathcal{N}_{\mathrm{sub}}\\ j\ge i}}
        p_{i,j}x_{k,i}x_{k,j}
        +
        \sum_{i\in\mathcal{N}_{\mathrm{sub}}}
        u_i^{(k)}x_{k,i}
    \right].
\end{align}
Terms involving only fixed variables are constants and are omitted from the subproblem QUBO.

We next define the constraints of the subproblem.
For each selected knapsack $k\in\mathcal{K}_{\mathrm{sub}}$, the fixed items in $\mathcal{N}_{\mathrm{fix}}^{(m)}(k)$ already use part of the capacity.
Thus, the residual capacity available to the free items is
\begin{align}
    \label{eq: QMKP_residual_capacity}
    C'_k
    =
    C_k
    -
    \sum_{j\in\mathcal{N}_{\mathrm{fix}}^{(m)}(k)} w_j.
\end{align}
Since the current solution is feasible, $C'_k$ is nonnegative.
The capacity constraint in the subproblem is then
\begin{align}
    \sum_{i\in\mathcal{N}_{\mathrm{sub}}}w_i x_{k,i}\leq C'_k,
    \quad \forall k\in\mathcal{K}_{\mathrm{sub}}.
\end{align}
The at-most-one-knapsack constraint for the selected items is
\begin{align}
    \sum_{k\in\mathcal{K}_{\mathrm{sub}}}x_{k,i}\leq 1,
    \quad \forall i\in\mathcal{N}_{\mathrm{sub}}.
\end{align}

Using the residual capacity, the number of auxiliary variables for knapsack $k$ in the subproblem is
\begin{align}
    D'_k
    =
    \left\lceil \log_2(C'_k+1)\right\rceil.
\end{align}
The QUBO energy function for the subproblem is
\begin{align}
    \label{eq: QMKPenergy_sub}
    {\mathcal{H}'}^{\mathrm{QMKP}}(\boldsymbol{x},\boldsymbol{y})
    &=
    {\mathcal{H}'}^{\mathrm{QMKP}}_{\mathrm{obj}}(\boldsymbol{x})
    \nonumber\\
    &+
    \mu_\mathrm{E}
    \sum_{k\in\mathcal{K}_{\mathrm{sub}}}
    \Bigg[
        \sum_{d=0}^{D'_k-1}2^d y_{k,d}
        -
        \left(2^{D'_k}-1-C'_k\right)\nonumber\\
        &\qquad\qquad\qquad\qquad\quad-
        \sum_{i\in\mathcal{N}_{\mathrm{sub}}}w_i x_{k,i}
    \Bigg]^2
    \nonumber\\
    &+
    \mu_\mathrm{F}
    \sum_{i\in\mathcal{N}_{\mathrm{sub}}}
    \left(
        \sum_{k\in\mathcal{K}_{\mathrm{sub}}}x_{k,i}
    \right)
    \left(
        \sum_{k\in\mathcal{K}_{\mathrm{sub}}}x_{k,i}-1
    \right).
\end{align}

This subproblem generation method preserves feasibility in the following sense.
Consider any feasible current solution and any feasible solution of the subproblem.
For each selected knapsack, the free items use at most the residual capacity $C'_k$, and the fixed items use at most the remaining capacity $C_k-C'_k$.
For each unselected knapsack, the contents are unchanged from the feasible current solution.
Therefore, constraint E is satisfied after reinsertion.
For constraint F, each selected item is assigned to at most one selected knapsack, and its variables for the unselected knapsacks are fixed to zero because the item was not placed in any unselected knapsack in the current solution.
The assignments of all other items are unchanged.
Therefore, constraint F is also satisfied after reinsertion.
Consequently, reinserting any feasible subproblem solution yields a feasible solution of the original QMKP.
When no feasible subproblem solution is obtained, the current solution is kept unchanged, as described in Sec.~\ref{subsec:LNS}.

The number of binary variables in the subproblem is
\begin{align}
    \label{eq: QMKP_sub_variable_number}
    n_{\mathrm{var}}^{\mathrm{QMKP}}
    =
    N_{\mathrm{sel}}K'
    +
    \sum_{k\in\mathcal{K}_{\mathrm{sub}}}D'_k.
\end{align}
When $N_{\mathrm{sel}}=N'$, this becomes
\begin{align}
    n_{\mathrm{var}}^{\mathrm{QMKP}}
    =
    N'K'
    +
    \sum_{k\in\mathcal{K}_{\mathrm{sub}}}D'_k.
\end{align}
Thus, the subproblem size can be controlled through the two parameters $K'$ and $N'$ independently of the original problem size, although the auxiliary-variable term also depends on the residual capacities.

\subsection{Problem Setting}
\label{subsec: QMKP_setting}

Table~\ref{tab: QMKPsettings} shows the QMKP instance and the subproblem-generation parameters used in the numerical experiments.
\begin{table}
  \centering
  \setlength{\abovecaptionskip}{4pt}
  \setlength{\belowcaptionskip}{6pt}
  \caption{Settings of the target QMKP and the subproblem-generation parameters.}
  \label{tab: QMKPsettings}
  \begin{tabular}{p{4cm} p{3cm}}
      \toprule
      \parbox[c]{4cm}{\centering Parameter}
      & \parbox[c]{3cm}{\centering Value} \\
      \hline
      \parbox[c]{4cm}{\centering $N$: Number of items}
      & \parbox[c]{3cm}{\centering 200} \\
      \parbox[c]{4cm}{\centering $K$: Number of knapsacks}
      & \parbox[c]{3cm}{\centering 5} \\
      \parbox[c]{4cm}{\centering $p_{i,j}$, $w_i$: Values and weights of the items}
      & \parbox[c]{3cm}{\centering QKP instance ``jeu\_200\_75\_1''~\cite{SoutifQKP}} \\
      \parbox[c]{4cm}{\centering $C_k$: Capacity of knapsack $k$}
      & \parbox[c]{3cm}{\centering $\left\lfloor 0.8\sum_{i\in\mathcal{N}}w_i/K \right\rfloor$~\cite{hiley2006quadratic}} \\
      \hline
      \parbox[c]{4cm}{\centering $K'$: Number of knapsacks in the subproblem}
      & \parbox[c]{3cm}{\centering 2, 3, 4, 5} \\
      \parbox[c]{4cm}{\centering $N'$: Prescribed number of items in the subproblem}
      & \parbox[c]{3cm}{\centering 10, 20, 30, 40} \\
      \bottomrule
  \end{tabular}
\end{table}

The item values, correlated values, and weights were taken from the QKP benchmark instance ``jeu\_200\_75\_1''~\cite{SoutifQKP}.
Following Ref.~\cite{hiley2006quadratic}, we generated a QMKP instance by specifying the number of knapsacks and assigning the same capacity to all knapsacks.
Although the subproblem can also be defined for $K'=1$, we used $K'\geq 2$ in the numerical experiments because we focus on updates involving reassignment among multiple knapsacks.

For each pair of $K'$ and $N'$, we performed 10 independent runs and evaluated the solution quality using the average total profit.
The error bars in Fig.~\ref{fig: QMKP_sub_variables} represent the standard deviations over these 10 runs.
Each LNS run was performed for 100 iterations.
The computation time for each iteration was set to 10~s.
The penalty coefficients were set to $\mu_\mathrm{E}=N$ and $\mu_\mathrm{F}=100N$ in all settings.
These values were selected from preliminary trials because they yielded feasible solutions in the tested settings.
After each run, the feasibility of the obtained solution was checked.

The selected knapsacks and items were sampled uniformly at random according to the procedure described in Sec.~\ref{subsec: QMKP_proposed}.
As in the VRP experiments, a feasible solution obtained by the naive method was used as the common initial solution for all QMKP-LNS settings.

\subsection{Results for the QMKP}
\label{subsec: QMKP_result}

\begin{figure}[t]
  \centering
  \includegraphics[width=0.95\linewidth]{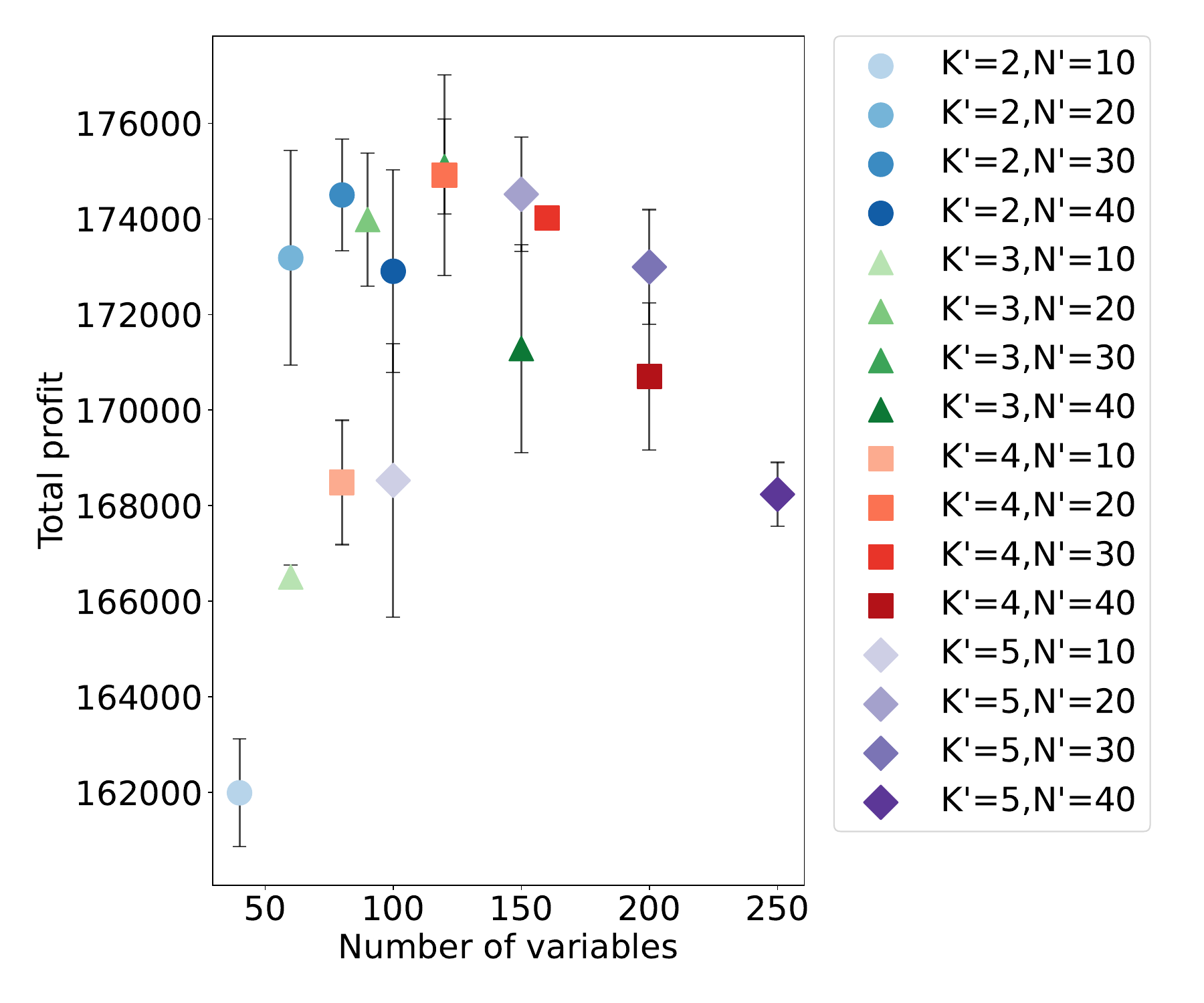}
  \caption{Relation between the number of binary variables included in the subproblem and the total profit after 100 LNS iterations for the QMKP with 200 items and 5 knapsacks. The points represent the average values over 10 runs, and the error bars represent the standard deviations.}
  \label{fig: QMKP_sub_variables}
\end{figure}

Figure~\ref{fig: QMKP_sub_variables} shows the relation between the number of binary variables in the subproblem and the total profit after 100 LNS iterations.
Because the QMKP is a maximization problem, a larger total profit indicates a higher-quality solution.
The number of variables includes both the decision variables and the auxiliary variables.
However, according to Eq.~\eqref{eq: QMKP_sub_variable_number}, the number of auxiliary variables varies across iterations. 
Therefore, for convenience in Fig.~\ref{fig: QMKP_sub_variables}, the number of auxiliary variables is formally unified with that of the original problem, and the total number of variables is calculated as follows:
\begin{align}
    n_{\mathrm{var}}^{\mathrm{QMKP}}
    =
    N'K'
    +
    \sum_{k\in\mathcal{K}_{\mathrm{sub}}}D_k.
\end{align}
As previously mentioned, since the term $N'K'$ provides the dominant contribution, this evaluation does not deviate significantly from the actual number of variables.

We first compare the LNS results with the naive method, which directly solves the full QUBO of the QMKP without LNS-based reduction.
When the naive method was run 10 times with a computation time of 300 s per run, the average total profit was approximately 150000.
In contrast, several QMKP-LNS settings produced average total profits exceeding 170000 after 100 iterations.
This result shows that, under the tested settings, the LNS construction for the QMKP can improve the solution obtained by directly solving the full QUBO.
As in the VRP case, this comparison should be interpreted as a comparison under the stated computational budgets and should not be read as an end-to-end wall-clock-time comparison.

Among the tested QMKP-LNS settings, the largest average total profits were obtained when the number of variables in the subproblem was approximately 100 to 150.
This range is substantially smaller than the range observed for the VRP in Sec.~\ref{sec: result}.
This difference suggests that the appropriate subproblem size depends on the problem class, the QUBO formulation, and the Ising-machine settings.
When the number of variables deviated from this range, the average total profit tended to decrease.
The decrease was particularly clear for settings with fewer variables.

These observations are qualitatively consistent with the VRP results in the sense that very small and very large subproblems did not give the best solution quality in the tested settings.
At the same time, the appropriate range of the number of variables differed substantially between the VRP and the QMKP.
Together with the comparison between LNS-V and LNS-VT for the VRP, the QMKP results support the view that feasibility preservation alone is not sufficient for designing effective subproblems in Ising-machine-assisted LNS.
They also support the importance of adjusting the subproblem size independently of the problem size and structure.

\section{Conclusion}
\label{sec: conclusion}

In this study, we investigated Ising-machine-assisted LNS for large-scale constrained combinatorial optimization problems.
For the VRP, we proposed LNS-VT, a subproblem generation method that introduces the number of consecutive steps re-optimized for each selected vehicle as an additional control parameter.
The existing method, LNS-V, preserves feasibility, but its subproblem size is controlled only by the number of selected vehicles and depends strongly on the number of steps of the original problem.
In contrast, LNS-VT allows the number of variables in the subproblem to be tuned more finely while preserving feasibility.

We evaluated LNS-VT through numerical experiments on a VRP with 300 non-depot sites and 5 vehicles.
Under the tested settings, the best LNS-VT setting reduced the average objective value by approximately 10\% relative to LNS-V after 100 iterations.
It also reached a solution quality comparable to the final quality of LNS-V with approximately 30\% of the iterations.
This comparison is based on the number of LNS iterations under the same computation time per subproblem QUBO and does not by itself establish an end-to-end wall-clock-time advantage.
We also observed that the appropriate subproblem setting changed with the quality of the current solution.
This result suggests that the subproblem size should be selected with the stage of the search in mind.

We further applied the same design principle to the QMKP.
For the QMKP, we constructed a feasibility-preserving subproblem generation method in which the number of selected knapsacks and the number of selected items control the subproblem size.
In the tested QMKP instance, several QMKP-LNS settings produced higher average total profits than the naive method of directly solving the full QUBO.
The appropriate range of the number of variables in the subproblem was approximately 100 to 150, which was substantially smaller than the range observed for the VRP.
These results suggest that the appropriate subproblem size depends on the problem class, the QUBO formulation, and the Ising-machine settings.

Overall, the results of this study support the view that feasibility preservation alone is not sufficient for designing effective subproblems in Ising-machine-assisted LNS.
Controlling the number of variables in the subproblem is also an important design factor.
At the same time, the present conclusions are based on the tested VRP and QMKP instances and on the Ising-machine settings used in this study.
Further experiments are needed to determine how broadly the observed tendencies hold.

Future work includes improving the selection rule for the components included in each subproblem.
In the present LNS-VT implementation for the VRP, the selected vehicles and extraction intervals are sampled uniformly at random, and the target sites are determined by the current solution and the sampled intervals.
This random selection may repeatedly choose similar subproblems or choose route segments that are already locally well optimized.
A possible direction is to prioritize route segments with larger expected improvement, for example by using route crossings, spatial proximity, or past update history as selection criteria.

Another direction is to develop a method for estimating an appropriate subproblem size without performing an exhaustive parameter sweep.
In the present study, the appropriate number of variables was determined empirically by testing multiple parameter settings.
Analyzing the interaction structure of the QUBO, the constraint structure of the original problem, and the quality of the current solution may provide predictors of an appropriate subproblem size.
Such predictors would be useful for adaptive LNS strategies that change the subproblem-generation parameters during the search.

\section*{Acknowledgments}
The authors thank Taiga Hayashi and Mahiro Komaki for carefully reading the manuscript and providing valuable comments. 
The computations in this work were partially performed using the facilities of the Supercomputer Center, the Institute for Solid State Physics, The University of Tokyo.
This work was partially supported by the Japan Society for the Promotion of Science (JSPS) KAKENHI (Grant Number JP23H05447), the Council for Science, Technology, and Innovation (CSTI) through the Cross-ministerial Strategic Innovation Promotion Program (SIP), ``Promoting the application of advanced quantum technology platforms to social issues'' (Funding agency: QST), Japan Science and Technology Agency (JST) (Grant Number JPMJPF2221). S. Tanaka wishes to express gratitude to the World Premier International Research Center Initiative (WPI), MEXT, Japan, for supporting the Human Biology-Microbiome-Quantum Research Center (Bio2Q).

\bibliographystyle{jpsj}
\bibliography{reference}

\end{document}